\documentstyle[prb,twocolumn,psfig,aps]{revtex}
\begin{document}

\newcommand{\lb}{\langle}
\newcommand{\rb}{\rangle}
\renewcommand{\vec}[1]{{\bf #1}}
\newcommand{\ek}{\varepsilon(\vec{k})}
\newcommand{\kmsi}{_{\vec{k}-\sigma}}
\newcommand{\nmsi}{n_{-\sigma}}
\newcommand{\nsi}{n_{\sigma}}
\newcommand{\ksi}{_{\vec{k}\sigma}}
\newcommand{\isi}{_{i\sigma}}
\newcommand{\imsi}{_{i-\sigma}}
\newcommand{\jsi}{_{j\sigma}}
\newcommand{\jmsi}{_{j-\sigma}}
\newcommand{\llongrightarrow}
{-\!\!\!-\!\!\!-\!\!\!-\!\!\!\longrightarrow}
\newcommand{\arrowle}{
   \mbox{\raisebox{-0.5ex}{$\stackrel{<}{\rightarrow}$}}} 
\newcommand{\ix}[1]{{\textrm{\scriptsize #1}}}
\newcommand{\befig}[1]{\begin{figure}[#1]}
\newcommand{\eefig}{\end{figure}}

\addtolength{\oddsidemargin}{2ex}
\addtolength{\evensidemargin}{-9ex}
\addtolength{\textheight}{6ex}

\title{Magnetism in the single-band Hubbard model}
\author{T. Herrmann and W. Nolting}
\address{Humboldt-Universit\"at zu Berlin, Institut f\"ur Physik, Invalidenstr.\ 110, 10115 Berlin, Germany}
\date{2.Sep.96; revised version 7.Jan.97}
\maketitle

\begin{abstract}
A self-consistent spectral density approach (SDA) is applied to the Hubbard
model  to investigate the possibility of spontaneous ferro- and
antiferromagnetism. Starting point is a two-pole ansatz for the single-electron
spectral density, the free parameter of which can be interpreted as  energies
and spectral weights of respective quasiparticle excitations. They are
determined by fitting exactly calculated spectral moments. The resulting
self-energy consists of a local  and a non-local part. The higher correlation
functions entering the spin-dependent  local part can be expressed  as
functionals of the  single-electron spectral density. Under certain conditions
for the decisive model parameters (Coulomb interaction $U$, Bloch-bandwidth $W$,
band occupation $n$, temperature $T$) the local part of the self-energy  gives
rise to a spin-dependent band shift, thus allowing  for spontaneous band
magnetism. As a function of temperature, second order phase transitions 
are found away from half filling, but
close to half filling the system exhibits a tendency towards 
first order transitions.
The non-local self-energy part is determined by use of proper
two-particle spectral densities. Its main influence concerns a (possibly
spin-dependent) narrowing of the quasiparticle bands with the tendency to stabilize
magnetic solutions. The non-local self-energy part disappears in the limit of
infinite dimensions. We present a full evaluation of the Hubbard model in terms
of quasiparticle densities of states, quasiparticle dispersions, magnetic phase
diagram, critical temperatures ($T_{C}$, $T_{N}$) as well as spin and particle
correlation functions. Special attention is focused on the non-locality of the
electronic self-energy, for which some rigorous limiting cases are worked out.
\end{abstract}


\section{INTRODUCTION}
Important consequences of strong correlations in narrow energy bands as band
magnetism, metal-insulator transitions or high-$T_{C}$ superconductivity are
normally investigated by rather simplified theoretical models. Of special
interest is the Hubbard model \cite{Hub63} which describes itinerant electrons
in a single non-degenerate energy band interacting via an on-site Coulomb
interaction $U$.  It incorporates in the simplest way  the interplay of the
kinetic energy, the Coulomb interaction, the Pauli  principle and the
bandstructure and its consequences for the electronic and magnetic properties
of the band system.  Originally, the model was thought to explain the
bandmagnetism of the transition metals and their compounds  as well as  the
temperature, pressure  or doping driven metal-insulator  transition  of some
metal oxides. The latter were first  recognized by Mott \cite{Mot49} as
contradicting conventional band theory (Mott transition).  It is to the merit
of Hubbard \cite{Hub64b} to have demonstrated, how strong electron correlations
may cause  insulating behaviour of not fully occupied energy bands.

The in principle rather simple model provokes nevertheless a non-trivial
many body problem, that  could be solved up to now only for some special cases
as, e.g., the ground state of the one-dimensional system \cite{LW68}. Other
rigorous  statements  concern the possibility of  collective magnetic order  in
two and three-dimensional lattices \cite{Gho71,Nag66,Lie89}.  In the first years
after Hubbard's pioneering work, the investigation has mainly been focused on
the existence of ferro- and antiferromagnetic solutions of the  single-band model
\cite{Rot69,Nol72}, where, however,  approximations had to be tolerated when
tackling the respective  magnetic phase diagram.  The interest in the Hubbard
model has got a dramatic upsurge in the recent past  when its relevance to the
high temperature superconductivity  became manifest \cite{AS91}.  It has been 
recognized that in the limit  of infinite dimensions $d$ the electronic
self-energy is wave-vector  independent \cite{MV89,Mue89,Vol94}. That
simplifies but nevertheless does not solve the problem. Substantial progress
has been made by the observation \cite{JV92,Jar92,GK92} that for
$d\rightarrow\infty$ the self-energy of the Hubbard model and the self-energy of
the Anderson model are the same functionals of their respective local Green
functions. However, up to now one does not know the exact functional form. With
respect to the superconductivity problem, most of the recent investigations
have been done for the half filled or almost half filled energy band. 
Compared to this, relatively little effort is concentrated on the 
search for magnetic solutions of the Hubbard model. Some interesting 
recent investigations which address the magnetic behaviour  
can be found in \cite{MHHH93,Uhr96,Ulm96,SSST89,FJ95}.

Ferromagnetic  solutions are to be expected, if at all, only in the strong coupling
regime $U/W>1$ (W: Bloch-bandwidth). It has been demonstrated \cite{Nol72,NB89}
that such strong electron correlations are reasonable well accounted for by the
"{\em spectral density approach}" (SDA) which basically consists in a two-pole
ansatz for the single-electron spectral density. The free parameters in 
the two-pole spectral density are
self-consistently fixed by equating exactly calculated spectral moments. The
advantages of the SDA rest on the clear and simple concept and its
non-perturbative character. It turns out to be essentially equivalent to the
Roth method \cite{Rot69,BE95} and to the Mori-projector formalism
\cite{Mor65b,Zwa61}. Various applications to Bose-, Fermi-, and classical
systems \cite{NO73,CCeal84,BBR86} have proven the efficiency of the SDA.
Previous applications of the SDA to the problem of band magnetism in the
Hubbard model for the bulk \cite{NB89} as well as for systems with reduced
symmetry \cite{PN96} have been restricted to a local self-energy that could be
derived  self-consistently. The wave-vector dependent part of the self-energy,
being built up by higher correlation functions referring to double hopping,
spinflip and density correlations, has been neglected. Recently, the
justification of such a neglect has been questioned \cite{BE95,EOMS94} and
investigated in detail for the paramagnetic two-dimensional system. 

In this paper we present for the first time a full evaluation of the SDA to the
Hubbard model with respect to the possibility of spontaneous ferro- and
antiferromagnetism~\cite{foot1}. Special attention is devoted to the question to what extent
the non-locality of the electronic self-energy influences the stability of
magnetic order. For this purpose we have organized the paper as follows. In
sec.\,\ref{sec_many} we give a short introduction to the underlying many
body problem. The theory of the spectral density approach to the Hubbard model
is developed in detail in Sec.\,\ref{sec_sda}.  In Sec.\,\ref{sec_limits} we
derive some analytic expressions for the case of half-filling ($n=1$) and  in the
limit of strong Coulomb interaction $U\rightarrow\infty$. The results of the
numerical calculations are discussed in Sec.\,\ref{disc}, where we investigate
para-, ferro- as well as antiferromagnetic solutions. We discuss the properties
of the system by means of a magnetic phase diagram, magnetization curves,
critical temperatures, quasiparticle density of states and quasiparticle
dispersions. In Sec.\,\ref{sec_sum} we give a short summary and some concluding
remarks.
\section{MANY BODY PROBLEM}\label{sec_many}
Starting point of our investigation is the single-band Hubbard model:
\begin{equation}  \label{hubbardhamilton}
	{\cal H}=
	\sum_{i,j,\sigma}(T_{ij}-\mu\delta_{ij}) 
	c_{i\sigma}^{\dagger} c_{j\sigma} \,\, 
	+\,\,\frac{U}{2} \sum_{i,\sigma} n_{i\sigma} n_{i-\sigma}.
\end{equation}  
$c_{i\sigma}^{\dagger}$ ($c_{j\sigma}$) is the creation (annihilation) operator
of an electron with spin $\sigma=\uparrow,\downarrow$ in a Wannier state at
lattice site $\vec{R}_{i}$. 
$n_{i\sigma}=c_{i\sigma}^{\dagger}c_{i\sigma}$ is the number operator and $U$
denotes the intraatomic Coulomb matrix element.  The intersite hopping
integrals
$T_{ij}$ are connected via Fourier transformation to the Bloch energies:
\begin{equation} \label{bdos}
	\varepsilon(\vec{k})=\frac{1}{N}\sum_{i,j}T_{ij} 
			  e^{-i\vec{k}(\vec{R}_{i}-\vec{R}_{j})}.
\end{equation}  
N is the number of lattice sites. The center of gravity of the 
Bloch band is given by:
\begin{equation}	
	T_{0}=\frac{1}{N}\sum_{\vec{k}}\varepsilon(\vec{k})=T_{ii}.
\end{equation}
Since the theory presented below does at this point not depend on the 
functional form of $\varepsilon(\vec{k})$ we do not specify the 
underlying lattice. In sec.~5 we will discuss some numerical results 
for the bcc-lattice and the square lattice. 
Other lattice structures, such as sc, fcc, hc-$d\!\!=\!\!\infty$,
fcc-$d\!\!=\!\!\infty$ will be considered in a forthcoming paper \cite{HN97}.

All interesting single-particle properties of the system are determined by the
retarded single-electron Green function:
\begin{eqnarray} 		
  	G_{\vec{k}\sigma}(E) &=& 
  	\lb\!\lb  c_{\vec{k}\sigma};c_{\vec{k}\sigma}^{\dagger}\rb\!\rb_{E}
  	=\frac{1}{N}\sum_{i,j} 
	 e^{-i\vec{k}\cdot(\vec{R}_{i}-\vec{R}_{j})}G_{ij\sigma}(E),
	 \nonumber\\
	G_{ij\sigma}(E) &=&
	\lb\!\lb  c_{i\sigma};c_{j\sigma}^{\dagger}\rb\!\rb_{E}=
	     \nonumber\\
	& &-i\int \limits_{0}^{+\infty}\!dt e^{-\frac{i}{\hbar}Et}
	\lb\left[c_{i\sigma}(t),c_{j\sigma}^{\dagger}(0)\right]_{+}\rb .
      \label{green_function}	
\end{eqnarray}
Here, $\vec{k}$ is a wave-vector of the first Brillouin zone and
$c_{\vec{k}\sigma}$  the Fourier transform of $c_{i\sigma}$. $[..,..]_{(+)-}$
denotes the (anti)commutator and $\lb ...\rb$ the grand-canonical average.

By introducing  the electronic self-energy $\Sigma_{\vec{k}\sigma}(E)$ via
the definition
\begin{equation}
	\lb\!\lb\left[c_{\vec{k}\sigma},
	\frac{1}{2}U\sum_{i\sigma}n_{i\sigma} n_{i-\sigma}\right]_{-};
	c_{\vec{k}\sigma}^{\dagger}\rb\!\rb_{E}
	=\Sigma_{\vec{k}\sigma}(E)G_{\vec{k}\sigma}(E),
\end{equation}
the equation of motion of the Green function is formally solved:
\begin{equation}	
	G_{\vec{k}\sigma}(E)=\frac{\hbar}{E-(\varepsilon(\vec{k})-\mu)
	-\Sigma_{\vec{k}\sigma}(E)}.
\end{equation}
The self-energy $\Sigma_{\vec{k}\sigma}(E)$ therefore gathers  all 
influences of the two particle interaction of the Hubbard model.

The approach to the Hubbard model in this paper is attached to the 
single-electron
spectral density
\begin{equation}	
	S_{\vec{k}\sigma}(E)=-\frac{1}{\pi}\mbox{Im}G_{\vec{k}\sigma}(E),
\end{equation}
which contains exactly the same information as the Green function or the
self-energy. Via (inverse) photoemission the spectral density is directly
related to  the experiment. 

By use of the spectral theorem one can  determine from the spectral density 
the average occupation number:
\begin{eqnarray}
	\lb n_{i\sigma}\rb =\lb c_{i\sigma}^{\dagger}c_{i\sigma}\rb&=&
	\int\limits_{-\infty}^{+\infty}\!dE f_{-}(E)
	S_{ii\sigma}(E-\mu)\\
	&=&\int\limits_{-\infty}^{+\infty}\!dE f_{-}(E)
	\rho_{\sigma}(E).
\end{eqnarray}
Here we have introduced the Fermi function $f_{-}(E)$ and the quasi-particle
density of states (QDOS):
\begin{equation}	
	\rho_{\sigma}(E)
	=\frac{1}{\hbar}S_{ii\sigma}(E-\mu)=
	\frac{1}{\hbar N}\sum_{\vec{k}}S_{\vec{k}\sigma}(E-\mu).
\end{equation}
If we  assume  translational symmetry, the  average occupation number
does not depend on the lattice site $\vec{R}_{i}$:
\begin{equation}
n_{\sigma}=\lb n_{i\sigma}\rb.
\end{equation}
The total occupation number is given by $n=n_{\uparrow}+n_{\downarrow}$.

A lot of  information about the spectral density can be drawn from its spectral
decomposition (Lehmann representation)
\begin{eqnarray}
	S_{\vec{k}\sigma}(E)&=&	   
	   \frac{\hbar}{\Xi}\sum_{n,m}
	       |\lb E_{n}|c_{\vec{k}\sigma}^{\dagger}|E_{m}\rb |^{2}              
             e^{-\beta E_{n}}\,(e^{\beta E}+1) 
         \nonumber \\& &    \times
         \delta[E-(E_{n}-E_{m})],
         \label{spectral_decomposition}
\end{eqnarray} 
where $\Xi$ is the grand canonical partitition function, $|E_{m}\rb$ an
$N$-particle, and $|E_{n}\rb$ an $(N+1)$-particle eigenstate  of the
Hamiltonian. Therefore, the energy differences $E_{n}-E_{m}$ correspond 
to the excitation energies required for adding a single electron to the
$N$-particle system.

Central quantities of our procedure are the moments 
$M_{\vec{k}\sigma}^{(n)}$ of the
spectral density $S_{\vec{k}\sigma}(E)$, for which two  equivalent 
representations exist.
One  is the usual definition  via an energy integral
\begin{equation}\label{moment1} 
	M_{\vec{k}\sigma}^{(n)}=\int\limits_{-\infty}^{+\infty}\!dE\, E^{n}
	S_{\vec{k}\sigma}(E),
\end{equation}
while the other involves only commutator relations between the 
construction operators and the
Hamiltonian and is, therefore, independent of the spectral density itself:
\begin{eqnarray}\label{moment2}	
	M_{\vec{k}\sigma}^{(n)}&=&\frac{1}{N}\sum_{i,j}e^{-i\vec{k}\cdot
	(\vec{R}_{i}-\vec{R}_{j})}
	\nonumber\\
	& &\hspace{-3ex}\times\lb \Bigg[\underbrace{\bigg[..
	\Big[c_{i\sigma}{\cal H}\Big]_{\!-}.. ,
            {\cal H}\bigg]_{\!-}}_{(n-p)-\textrm{times}},
             \underbrace{\bigg[{\cal H},..
             [{\cal H},c_{j\sigma}^{\dagger}]_{-}..
             \bigg]_{\!-}}_{p-\textrm{times}}\Bigg]_{\!+} \!\!\rb.                   
             \label{momente2}                
\end{eqnarray}  
p is an integer between $0$ and $n$ with $n=0,\,1,\,2, \dots$.
\\

Besides systems with translational symmetry we also want to investigate
antiferromagnetic structures. 
For simplicity we restrict the analytical calculations presented in the next
sections to the case of paramagnetism and ferromagnetism, respectively,
and give only a brief idea of the extension to
antiferromagnetism. The following is meant to introduce the notation
necessary to describe antiferromagnetic configurations.

First  we have to decompose the chemical
lattice into two equivalent magnetic sublattices $A$ and $B$. 
The original, chemical
lattice can then be described by a so-called magnetic Bravais lattice
($\vec{R}_{i}$)  with a two-atomic basis ($\vec{r}_{\alpha}$):
\begin{equation}
	\vec{R}_{i\alpha}=\vec{R}_{i}+\vec{r}_{\alpha},
	\quad(i=1,...,N/2;\,\alpha=A,B)
\end{equation}	
This new labeling of the lattice sites applies, of course, for the creation and
annihilation operators $c_{i\alpha\sigma}^{\dagger}$, 
$c_{i\alpha\sigma}$ as well. 
	                          
If we assume translational symmetry inside 
the magnetic sublattices, 
expectation values do not depend on the lattice site $\vec{R}_{i}$.
For example we write for the sublattice occupation number:
\begin{equation}
	n_{\alpha\sigma}=\lb n_{i\alpha\sigma}\rb=
	\lb c_{i\alpha\sigma}^{\dagger}c_{i\alpha\sigma}\rb .
\end{equation}

In the case of antiferromagnetic order, the dependence on 
the sublattice index $\alpha$ becomes important and 
$n_{A\sigma}=n_{B-\sigma}$ holds. This yields for the sublattice magnetization:
\begin{equation}
	m_{A}=n_{A\uparrow}-n_{A\downarrow}=
	      n_{B\downarrow}-n_{B\uparrow}=-m_{B}.
\end{equation}	
Paramagnetism and ferromagnetism can still be described by setting 
$n_{A\sigma}=n_{B\sigma}$.

All  calculations are carried out in wave-vector space. The actual
choice of the antiferromagnetic configuration, therefore, comes 
in through the single-electron energies $\varepsilon_{\alpha\beta}(\vec{k})$
only:
\begin{equation}
	\varepsilon_{\alpha\beta}(\vec{k})=\frac{2}{N}\sum_{i,j} 
	T_{ij}^{\alpha\beta}
	e^{-i\vec{k}\cdot(\vec{R}_{i}-\vec{R}_{j})}.\label{eps_afm}
\end{equation}

\section{SPECTRAL DENSITY APPROACH}\label{sec_sda}
We use a self-consistent spectral density approach (SDA) \cite{Nol72} 
to find an approximate
solution of the Hubbard Hamilton operator. The method is based on a physically
motivated ansatz for the single-electron spectral density. The  spectral density 
approach  has proven to be
very successful studying various many body problems such as Bose, Fermi and
classical systems \cite{NO73,CCeal84,BBR86}. The main advantages of this method
are the physically simple concept and the explicit non-perturbative character.
Recent applications use the SDA successfully for the 
investigation of the attractive
Hubbard model ($U<0$) \cite{SPR96}, the t-J-model \cite{Mas93} and  
for studying systems
with reduced dimension \cite{PN96}, surface
magnetism \cite{PN96b} and magnetism of thin films \cite{PN96c}. 
The SDA is believed to be a
good starting point for the investigation of high temperature
superconductivity \cite{BE95,MEHM95}. 

The work presented here, is mainly concerned with the
magnetic properties of the Hubbard model. For the first time we give a complete
evaluation of the SDA theory.
The SDA can be divided into three major steps:
\begin{itemize}
\item[(i)] The crucial point is to find a physically reasonable ansatz for the
spectral density. Some hints can be drawn, for example, from exactly known
limiting cases, sum rules for the peaks of the spectral density, spectral
decompositions, etc.
\item[(ii)] the mathematical ansatz in (i) will contain some free parameters,
which can be fitted by the exactly calculated  moments 
$M_{\vec{k}\sigma}^{(n)}$ of the spectral density. All
correlation functions, however, which occur in this procedure, have to be
re-expressed by the spectral density.
\item[(iii)] With (i) and (ii) one obtains  a closed set of equations, 
which can be solved self-consistently. 
\end{itemize}

If at all, ferromagnetism in the Hubbard model is to be expected in the strong
correlation limit $U > W$. Therefore, our ansatz for the spectral density
should be motivated in this limit. In the strict $W=0$ case, where no hopping
is allowed between the lattice sites, the spectral density  consists
of two weighted $\delta$-functions. The two $\delta$-peaks 
are located
at the energies $T_{0}-\mu$ and $T_{0}+U-\mu$, corresponding to the excitation
energies required to add a 
$(\sigma)$-electron to a lattice site
where a $(-\sigma)$-electron is already present or not.  

The interesting question is what happens, if
one allows for a small but non-zero hopping? Following Harris and Lange
\cite{HL67}, four effects are to be expected: A broadening of the 
$\delta$-peaks,  a shift of the center of gravity of the peaks, a rearrangement
of spectral weight  between the  peaks and, finally, the appearance of 
new, so-called 
"satellite" peaks  near the energies 
$E_{p\cdot U}=T_{0}+p\cdot U-\mu$ (with
$p\in Z\setminus\{0,1\}$). The first three effects can 
already be seen in a two site
Hubbard model. Harris and Lange have shown \cite{HL67}, that the two satellite
peaks at $E_{-U}$ and $E_{2\cdot U}$, which are closest to the main peaks,
have weight factors of order $(W/U)^{4}$, being therefore negligible for the 
strongly correlated system ($U\gg W$). 
The weights of the other satellite peaks
are even smaller. Therefore, we can conclude that, in the case 
of strong correlation the spectral density has a two peak structure. 
Besides the Kondo-like peak, which is beyond the scope of our present investigation
\cite{foot2}, this decisive point
in our theory is confirmed, for example, by recent
calculations in the limit of infinite lattice dimensions 
$d=\infty$ \cite{Jar92,JP93}.

If we neglect
quasiparticle damping effects, the following two-pole ansatz for the
single-electron spectral 
density is, therefore, physically reasonable:
\begin{equation}
	S_{\vec{k}\sigma}(E)=\hbar\sum_{j=1,2}\alpha_{j\sigma}(\vec{k})
	\delta(E-E_{j\sigma}(\vec{k})+\mu).\label{ansatz}
\end{equation}
The four free parameters $E_{1\sigma}(\vec{k})$, $E_{2\sigma}(\vec{k})$,
$\alpha_{1\sigma}(\vec{k})$, $\alpha_{2\sigma}(\vec{k})$ contained in the
ansatz (\ref{ansatz}) can be calculated using the two independent
representations (\ref{moment1}) and (\ref{moment2}) of the spectral moments.
For this we have to calculate the first four moments 
$M_{\vec{k}\sigma}^{(0)}$-$M_{\vec{k}\sigma}^{(3)}$, which are given, for
example, in ref.~\cite{Nol72}. After a straightforward calculation we 
finally get:
\begin{eqnarray}
	E_{(1,2)\sigma}(\vec{k})&=&\frac{1}{2} \bigg[ B_{\vec{k}-\sigma}+U+\ek
	                      \nonumber \\* 
	& &\hspace{-12ex} \mp \sqrt{\left(B\kmsi+U-\ek \right)^{2}+
	4U\nmsi \left( \ek-B\kmsi \right) } \: \bigg] 
				\label{sda_energiepole} , \\[0.3cm]
	\alpha_{1\sigma}(\vec{k})&=&
	\frac{E_{2\sigma}(\vec{k})-\ek-U n_{-\sigma}}
	     {E_{2\sigma}-E_{1\sigma}}
	                  \label{sda_gewicht1} , \\*[0.25cm]
	\alpha_{2\sigma}(\vec{k})&=&
	\frac{U n_{-\sigma}+\ek-E_{1\sigma}(\vec{k})}
	     {E_{2\sigma}-E_{1\sigma}}=1-\alpha_{1\sigma}(\vec{k})
	                  \label{sda_gewicht2}.	
\end{eqnarray}
Here, we introduced the correction term $B_{\vec{k}-\sigma}$, which contains
higher correlation functions (see below). The correction term $B_{\vec{k}-\sigma}$ 
turns out to be of decisive importance
with respect to the possibility of spontaneous magnetic order.
We want to emphasize, that (\ref{sda_energiepole})-(\ref{sda_gewicht2})
reproduce the exact results obtained 
by Harris and Lange \cite{HL67} in the
strong correlation limit. 

It is also interesting to look at the self-energy 
$\Sigma\ksi^{SDA}(E)$, which is connected to the spectral density via:
\begin{equation}
	S_{\vec{k}\sigma}(E)=
	\hbar\delta(E-\Sigma_{\vec{k}\sigma}(E)+\mu).\label{selbst_ansatz}
\end{equation}
Using (\ref{sda_energiepole})-(\ref{sda_gewicht2}) we find:
\begin{equation}
	\label{sda_self_energy}
	\Sigma\ksi^{SDA}(E)=Un_{-\sigma}
	\frac{E+\mu-B\kmsi}{E+\mu-B\kmsi-U\left(1-n_{-\sigma}\right)}.
\end{equation}
Replacing $B_{\vec{k}-\sigma}$ in (\ref{sda_self_energy}) simply by the center
of gravity $T_{0}$ of the Bloch band,   reduces $\Sigma\ksi^{SDA}(E)$ 
to the so-called
Hubbard-I self-energy \cite{Hub63}. It is well known that the Hubbard-I
solution is not able to describe ferromagnetism. 
The main shortcoming of the Hubbard-I
solution with respect to the possibility of spontaneous magnetic order 
is the fact, that the centers of gravity of the quasiparticle subbands 
(so-called Hubbard bands) in the 
QDOS  are  spin-independent. 
Allowing  for such a spin-dependence via the correction
term $B_{\vec{k}-\sigma}$ is the decisive improvement of the SDA upon the
Hubbard-I solution.

The correction term $B_{\vec{k}-\sigma}$ can be split 
into a $\vec{k}$-independent and
a $\vec{k}$-dependent term:
\begin{equation}
	B\kmsi=B_{-\sigma}+F\kmsi.\label{bkms}
\end{equation}

Assuming translational invariance and hopping
between nearest neighbours only, the $\vec{k}$-dependence can be separated:
\begin{equation}
	F\kmsi=(\ek-T_{0})F_{-\sigma}.
\end{equation}
The remaining terms $B_{-\sigma}$ and $F_{-\sigma}$ are given by:
\begin{equation}
\label{bsigma}
	B_{-\sigma}\!-\!T_{0}=\frac{1}{\nmsi(1\!-\!\nmsi)}\frac{1}{N}
	\!\sum_{i,j}^{i\neq j} T_{ij}\lb c\imsi^{\dagger} 
	c\jmsi(2n\isi\!-1\!)\rb, 
\end{equation}	                 
\begin{eqnarray}	
	F_{-\sigma}&=& \frac{1}{\nmsi(1-\nmsi)} 
	[F_{-\sigma}^{(1)}+F_{-\sigma}^{(2)}+F_{-\sigma}^{(3)}]\\
	F_{-\sigma}^{(1)}&=&\lb n\imsi n\jmsi\rb -n_{-\sigma}^{2}
	\qquad\textrm{"density correlation",} 
	                             \\* 
	F_{-\sigma}^{(2)}&=&-\lb c\jsi^{\dagger} c\jmsi^{\dagger} 
	c\imsi c\isi\rb  
	\qquad\textrm{"double hopping cor.",} 
	                             \\* 
	F_{-\sigma}^{(3)}&=&-\lb c\jsi^{\dagger} c\imsi^{\dagger}
	c\jmsi c\isi\rb 
	\qquad\textrm{"spinflip correlation".} 
	\label{fksigma}
\end{eqnarray}
The so-called
spin dependent bandshift $B_{-\sigma}$ may lead to an energy shift in the 
spin-spectra, 
allowing, therefore, for an exchange splitting between the spin-up and
spin-down spectrum.

In former investigations of the magnetic properties of  the Hubbard model
within the SDA,  the bandwidth correction 
$F_{\vec{k}-\sigma}$ was  neglected and only the local part of the
self-energy was taken into account 
(see for example refs.\cite{NB89,PN96,SPR96,PN96b,PN96c,GN88,BdKNB90}). 
The bandwidth correction was identified to be of minor 
importance with respect to spontaneous magnetism, because 
the last two correlation functions $F^{(2)}_{-\sigma}$ and 
$F^{(3)}_{-\sigma}$ are effectively spin independent.
Further, the sum of $F_{\vec{k}-\sigma}$ over all 
wave-vectors of the first Brillouin
zone vanishes, so no additional exchange splitting is to be expected.
In the widely discussed case of infinite dimensions ($d=\infty$) the
self-energy becomes local, i.e., $F_{\vec{k}-\sigma}$ vanishes. 
Many approximations for the three-dimensional Hubbard model
also use a local self-energy for simplicity.
Nevertheless, in three and lower dimensions the spin and wave-vector 
dependent bandwidth correction 
$F_{\vec{k}-\sigma}$ can induce
substantial changes in the shape and the width of the QDOS. Since magnetism is
favoured by narrow energy bands, the bandwidth correction may  also 
have a significant
influence on the magnetic behaviour. 
One major part of this work is
concerned with the influence of the non-locality ($\vec{k}$-dependence) 
of the self-energy in three
and lower dimensions. 
In particular, the influence of these non-local terms
on the magnetic behaviour will be discussed. 

In one and two dimensions Beenen and Edwards \cite{BE95} and Mehlig et. al. 
\cite{MEHM95} showed for the case of paramagnetism,  that
the inclusion of the $\vec{k}$-dependence leads to a rather good agreement
of the quasiparticle dispersion calculated with the SDA and 
recent quantum Monte Carlo and exact diagonalization results
\cite{BSW94a,BSW94b}.

One of the more crude approximations in the ansatz (\ref{ansatz}) 
for the spectral density
is the fact, that quasiparticle damping is not included. The imaginary part of
the self-energy (\ref{sda_self_energy}), therefore, is identical to zero. 
Since the magnetic variables, like the magnetization  $m$, are given by an
energy integration over the spectral density, the broadening of the peaks in
the spectral density should not
be very important. However, one has to take this question seriously. 
In a recent paper \cite{HN96}, we investigated the influence of quasiparticle
damping on the magnetic stability by a proper combination of the SDA and the
coherent potential approximation. Guided by exact results in the strong
coupling limit, we proposed a modified alloy analogy, which allows for
spontaneous ferromagnetic solutions. 
Although in the paramagnetic case the results obtained by the modified alloy
analogy are almost identical to the SDA-predictions, 
the region where the system exhibits ferromagnetism
is  strongly reduced if quasiparticle damping is included \cite{HN96}.
Further work
on the influence of damping with respect to magnetism 
is in progress. 

What remains to be done in order to get a closed set of equations is to
express the two terms $B_{-\sigma}$ and $F_{-\sigma}$  by means of the
spectral density:
\\

\textbf{Spin-dependent bandshift:}
The higher correlation function 
$\lb n_{i\sigma}c_{i-\sigma}^{\dagger} c\jmsi\rb$ 
which appears in the spin-dependent bandshift
$B_{-\sigma}$ is exactly determined by the single-electron 
spectral density
\cite{GN88,NB89}: 
\begin{eqnarray}
	\lb n_{i\sigma}c_{i-\sigma}^{\dagger} c\jmsi\rb&=&
	\frac{1}{UN\hbar}\sum_{\vec{k}}
	e^{-i\vec{k}\cdot(\vec{R}_{i}-\vec{R}_{j})}
	\nonumber\\
	& &\hspace{-6ex}\times\int\limits_{-\infty}^{+\infty}\!\!dE f_{-}(E) 
	\left[E\!-\!\ek\right] S\kmsi(E\!-\!\mu).\label{nicjci}
\end{eqnarray}
This leads, together with the hopping correlation function 
$\lb c_{i-\sigma}^{\dagger} c\jmsi\rb$, which is given directly by the spectral
theorem
\begin{equation}	
	\lb c_{i-\sigma}^{\dagger} c\jmsi\rb=
	\frac{1}{N\hbar}\sum_{\vec{k}}e^{-i\vec{k}\cdot(\vec{R}_{i}-\vec{R}_{j})}
	\!\int\limits_{-\infty}^{+\infty}\!\!dE f_{-}(E) 
	S\kmsi(E\!-\!\mu),\label{hopping}
\end{equation} 
to the final expression for $B_{-\sigma}$:
\begin{eqnarray}
	B_{-\sigma}-T_{0}&=&
	\frac{1}{\nmsi(1-\nmsi)}\frac{1}{N}\sum_{\vec{k}} 
	\int\limits_{-\infty}^{+\infty}dE f_{-}(E)(\ek -T_{0})
	                   \nonumber \\
	& &\times
	\left[\frac{2}{U}\left[E-\ek\right]-1\right] S\kmsi(E-\mu).
	\label{sda_bandverschiebung}
\end{eqnarray}
\mbox{}\\

\textbf{Bandwidth correction:}
The evaluation of the higher correlation functions contained in the bandwidth
correction $F_{-\sigma}$ turns out to be more difficult. 
For simplicity we restrict the detailed calculation to the case
of the spinflip correlation function $F^{(3)}_{-\sigma}$. 
The evaluation of the
density and the double hopping correlation functions follows exactly
the same line.

First we rewrite $F^{(3)}_{-\sigma}$ as:
\begin{equation} \label{spinflip}
	F_{-\sigma}^{(3)}=-\sum_{l}\delta_{lj}
	\lb c_{l\sigma}^{\dagger} c_{j+\Delta-\sigma}^{\dagger} 
	 c\jmsi c_{j+\Delta\sigma}\rb,
\end{equation}
where $\Delta=\vec{R}_{i}-\vec{R}_{j}$ is a lattice vector which connects two
neighbouring lattice sites. In the following we assume translational invariance
and hopping only between nearest neighbours. Therefore, $F^{(3)}_{-\sigma}$
does not depend on the explicit value of $\Delta$.

If we now introduce the higher spectral density $A^{(3)}_{\vec{k}\sigma}$ via
\begin{eqnarray}
	A^{(3)}_{\vec{k}\sigma} &=&	       
	       \frac{1}{N}\sum_{i,j} 
	       \int \limits_{-\infty}^{+\infty}\!\!d(t\!-\!t')
	 	e^{-i\vec{k}\cdot(\vec{R}_{i}-\vec{R}_{j})}
	 	\nonumber\\
	 	& &\hspace{7ex}e^{\frac{i}{\hbar}E(t\!-\!t')}A_{jl\sigma}^{(3)}(t\!-\!t'),\\
	A_{jl\sigma}^{(3)}(t\!-\!t')&=&\frac{1}{2\pi}
	\lb \left[c_{j+\Delta-\sigma}^{\dagger}
	c\jmsi c_{j+\Delta\sigma}(t),
	       c_{l\sigma}^{\dagger}(t')\right]_{+}\rb, \label{def_ajlsigma}
\end{eqnarray}
it follows from the spectral theorem that $F^{(3)}_{-\sigma}$ is given by:
\begin{equation} \label{spinflip2}
	F_{-\sigma}^{(3)}=
	-\frac{1}{N}\sum_{\vec{k}}
	\int\limits_{-\infty}^{+\infty}dE 
	f_{-}(E)A_{\vec{k}\sigma}^{(3)}(E-\mu).
	\label{spinflip3}
\end{equation}       
To get further information about the structure of $A^{(3)}_{\vec{k}\sigma}$,
we look at its spectral decomposition:
\begin{eqnarray}
	A\ksi^{(3)}(E)&=& \frac{\hbar}{\Xi}\sum_{n,m}
	\lb E_{n}|\frac{1}{\sqrt{N}}\sum_{j}
	e^{i\vec{k}\vec{R}_{j}}c_{j+\Delta-\sigma}^{\dagger}
	c\jmsi c_{j+\Delta\sigma}|E_{m}\rb  
	\nonumber\\*
	& &\hspace{-8ex}\times \lb E_{m}|c_{\vec{k}\sigma}^{\dagger}|E_{n}\rb              
         e^{-\beta E_{n}}\,(e^{\beta E}+1)\,
          \delta[E-(E_{n}-E_{m})].\label{ajlsigma}
\end{eqnarray}
The matrix elements $\lb E_{n}|\dots|E_{m}\rb$ determine the weight of 
$A^{(3)}_{\vec{k}\sigma}(E)$ at the excitation energies $E=E_{n}-E_{m}$.
Therefore $A^{(3)}_{\vec{k}\sigma}(E)$ is non-zero only if the corresponding
matrix elements do not vanish. 
Since (\ref{ajlsigma}) and the spectral decomposition of 
$S_{\vec{k}\sigma}(E)$ (\ref{spectral_decomposition}) involve the same
matrix element $\lb E_{m}|c_{\vec{k}\sigma}^{\dagger}|E_{n}\rb$, 
the excitation energies $E_{n}-E_{m}$ are exactly the same as in
$S_{\vec{k}\sigma}(E)$. It is, therefore, 
reasonable to assume that the positions of the quasiparticle peaks
in $A^{(3)}_{\vec{k}\sigma}(E)$ are  also  the same  
as in  $S_{\vec{k}\sigma}(E)$.
However, because the other matrix element in (\ref{ajlsigma}) is not equal to
the one in (\ref{spectral_decomposition}),
the distribution of spectral weight over the peaks will  be different 
in $A^{(3)}_{\vec{k}\sigma}(E)$ and $S_{\vec{k}\sigma}(E)$. 
We can, therefore, 
formulate an ansatz for 
$A^{(3)}_{\vec{k}\sigma}(E)$:
\begin{equation}
	A^{(3)}_{\vec{k}\sigma}(E)=\hbar\sum_{j=1,2}
	\beta^{(3)}_{j\sigma}(\vec{k})
	\delta(E-E_{j\sigma}(\vec{k})+\mu),\label{ansatz_ajlsigma}
\end{equation}
where  $E_{j\sigma}(\vec{k})$ ($j=1,2$) are the "old" positions 
of the peaks given by
(\ref{sda_energiepole})  and $\beta^{(3)}_{j\sigma}(\vec{k})$ the "new" weight
factors which have to be determined. As in the case of  
$S_{\vec{k}\sigma}(E)$ this can easily be done by evaluating the first two
spectral moments of $A^{(3)}_{\vec{k}\sigma}(E)$.
Note, that the same procedure yields the exact result (\ref{nicjci}) 
when applied to the
higher correlation function $\lb n_{i\sigma}c_{i-\sigma}^{\dagger} c\jmsi\rb$
involved in the bandshift $B_{-\sigma}$.

After a lengthy but straightforward calculation we find via 
(\ref{spinflip3}) the final expression for the three higher correlation
functions $F^{(1,2,3)}_{-\sigma}$:
\begin{eqnarray}
	F_{-\sigma}^{(1)}&=&-\frac{\eta_{-\sigma}
	\lb c_{i-\sigma}^{\dagger} c\jmsi\rb +\nu_{-\sigma}
	\lb n_{i\sigma}c_{i-\sigma}^{\dagger} c\jmsi\rb}
	{1+\nu_{0\sigma}\nu_{0-\sigma}},
				\label{density_corr}\\*	
	F_{-\sigma}^{(2)}&=&-\frac{(\eta_{\sigma}+\nu_{\sigma})
	\lb c_{i-\sigma}^{\dagger} c\jmsi\rb -\nu_{\sigma}
	\lb n_{i\sigma}c_{i-\sigma}^{\dagger} c\jmsi\rb }
	{1-\nu_{0\sigma}},	
			\label{double_corr}	\\*
	F_{-\sigma}^{(3)}&=&-\frac{\eta_{\sigma}
	\lb c_{i-\sigma}^{\dagger} c\jmsi\rb +\nu_{\sigma}
	\lb n_{i\sigma}c_{i-\sigma}^{\dagger} c\jmsi\rb }
	{1+\nu_{0\sigma}}.	\label{flip_corr}	 
\end{eqnarray}
For simplicity we introduced the abbreviations:
\begin{eqnarray}
	\eta_{\sigma}&=&\frac{1}{1-\nmsi}
	\left[\lb c_{i-\sigma}^{\dagger} c\jmsi\rb -
	\lb n_{i-\sigma}c_{i\sigma}^{\dagger} c\jsi\rb
	\right],  
			\label{musigma} 
			\\*
	\nu_{\sigma}&=&\frac{1}{\nmsi(1-\nmsi)} \left[
	\lb n_{i-\sigma}c_{i\sigma}^{\dagger} c\jsi\rb -
	\nmsi \lb c_{i\sigma}^{\dagger} c\jsi\rb \right],
			\label{nusigma} 
			\\*
	\nu_{0\sigma}&=&\frac{\lb n\isi n\imsi\rb -\nsi\nmsi}
	                     {\nmsi(1-\nmsi)}.
	            \label{nu0sigma}
\end{eqnarray}
Since $\lb n_{i-\sigma}c_{i\sigma}^{\dagger} c\jsi\rb$ and 
$\lb c_{i-\sigma}^{\dagger} c\jmsi\rb$ are given by (\ref{nicjci}) and 
(\ref{hopping}), respectively, the bandwidth correction $F_{-\sigma}$ can be
calculated by means of the single-electron spectral density 
$S_{\vec{k}\sigma}(E)$ only. 
Equations (\ref{ansatz})-(\ref{sda_gewicht2}), 	(\ref{bkms}), 
(\ref{nicjci})-(\ref{sda_bandverschiebung}) and
(\ref{density_corr})-(\ref{nu0sigma}) build a closed set of equations which can
be solved self-consistently.

Expressions (\ref{density_corr})-(\ref{nu0sigma}) for the bandwidth correction
are identical to the ones obtained by the method 
introduced by Roth \cite{Rot69}, both methods being, therefore, equivalent.

For the extension of the theory to antiferromagnetic systems, we use an
effective medium approach described in ref.~\cite{BdKNB90}. Since all further
calculations are strictly along the lines of \cite{BdKNB90} we do not  
give any details here. 
\\

\textbf{Paramagnetic static susceptibility:}
A lot of information about the magnetic behaviour of the system 
can  be drawn
from the static susceptibility:
\begin{equation}
	\label{def_chi}
	\chi(n,T,H)=\frac{1}{\mu_{0}}\left(\frac{\partial M}{\partial H}
	\right)_{T}.
\end{equation}
$\mu_{0}$ is the vacuum permeability, H denotes a static magnetic field and 
\begin{equation}
	M=\frac{N}{V}\mu_{B}(n_{\uparrow}-n_{\downarrow})
\end{equation}
is the magnetization of the system.

For studying  magnetic phase transitions we are interested  in the static
susceptibility in the limit $H\rightarrow 0$:
\begin{equation}
	\label{def_chi_0}
	\chi^{(0)}(n,T)=\frac{1}{\mu_{0}}\left(\frac{\partial M}{\partial H}
	\right)_{T}\Bigg|_{\textrm{paramag.}, H=0}.
\end{equation}
The poles of the paramagnetic static susceptibility $\chi^{(0)}(n,T)$ indicate
the  instabilities of the system against ferromagnetic order. Therefore, all
second order phase transitions from paramagnetism to ferromagnetism can be read
off from $\chi^{(0)}(n,T)$, which needs as an input only the paramagnetic 
solution.  
Besides this, the paramagnetic Curie temperature $\Theta$ as well as the
critical exponent $\gamma$ of the susceptibility can easily be obtained from
$\chi^{(0)}(n,T)$ \cite{NBdKB91}.

To calculate the static susceptibility we have to include a Zeeman term 
\begin{equation}
	H_{Z}=-\mu_{B} H (n_{i\uparrow}-n_{i\downarrow})
\end{equation}
in the Hamiltonian (\ref{hubbardhamilton}), which couples the magnetic field
$H$ to the local magnetic moment $m_{i}=n_{i\uparrow}-n_{i\downarrow}$. This
additional term does not alter or complicate the theory presented above.
However, when differentiating the magnetization with respect to the magnetic
field, we have to take into account that all expectation values are implicit
functions of $H$ \cite{NBdKB91}. All calculations can be done analytically,
whereas the results are rather lengthy.

\section{Limiting cases}\label{sec_limits}
Before we discuss the results of the numerical evaluation of the theory, 
we want to develop 
some analytical results for the case of  exact half filling $n=1$ and 
for the limit of strong Coulomb
interaction  $U\rightarrow\infty$.  Both terms, the bandshift $B_{-\sigma}$ and
the bandwidth correction $F_{-\sigma}$ simplify considerably in these cases.
Throughout this section we assume 
translational invariance, hopping
between nearest neigbours only and a symmetric BDOS. 

First we consider  the two correlation functions 
$\lb n_{i-\sigma}c_{i\sigma}^{\dagger} c\jsi\rb$  and 
$\lb c_{i-\sigma}^{\dagger} c\jmsi\rb$ which are given by (\ref{nicjci}) and 
(\ref{hopping}) respectively. 
The process described by $\lb n_{i-\sigma}c_{i\sigma}^{\dagger} c\jsi\rb$
can be understood as a hopping process of a  $(\sigma)$-electron with the
condition that one of the sites is occupied by a $(-\sigma)$-electron. Because
of the required double occupancy this process is  suppressed in the
limit of large $U$ and $n\le 1$. With (\ref{nicjci}) and 
(\ref{ansatz})-(\ref{sda_gewicht2}) we find indeed:
\begin{equation}
	\label{nicicjgrosseu}
	\lb c\imsi^{\dagger} c\jmsi n\isi\rb 
	\stackrel{n\le1,U\rightarrow\infty}{\llongrightarrow} 0.
\end{equation}	
For the half-filled band, paramagnetism and $T=0$K, a simple analytical
calculation involving (\ref{nicjci}), (\ref{hopping}) and
(\ref{ansatz})-(\ref{sda_gewicht2}) yields the relation:
\begin{equation}	
	\label{nicicjngleich1}
	\lb c\imsi^{\dagger} c\jmsi n\isi\rb=\frac{1}{2}
	\lb c_{i-\sigma}^{\dagger} c\jmsi\rb.
\end{equation}		
As $U\rightarrow\infty$ for the case of half filling  the hopping correlation 
$\lb c_{i-\sigma}^{\dagger} c\jmsi\rb$ should vanish,  because all hopping
processes require an energetically  unfavoured double occupancy. In fact,
(\ref{sda_energiepole})-(\ref{sda_gewicht2}) in (\ref{hopping}) give for 
($i\ne j$):
\begin{equation}
	\label{cicjgrosseu}
	\lb c\imsi^{\dagger} c\jmsi \rb 
	\stackrel{n=1,U\rightarrow\infty}{\llongrightarrow} 0.
\end{equation}	

We now want to focus on the spin-dependent 
bandshift $B_{-\sigma}$ and the bandwidth correction
$F_{-\sigma}$ in four different limiting cases:

\textbf{(i) Large $U$ and less than half filling:} With 
(\ref{nicicjgrosseu}) the bandshift $B_{-\sigma}$ is given by:
\begin{equation}
	B_{-\sigma}
	\stackrel{n<1,U\rightarrow\infty}{\llongrightarrow}
	-\frac{Z_{1}T_{1}}{n_{-\sigma}(1-n_{-\sigma})}
	\lb c_{i-\sigma}^{\dagger} c\jmsi\rb+T_{0}.
\end{equation}
$Z_{1}$ is the number of nearest neighbours and $T_{1}$ denotes the
hopping integral between neighbouring lattice sites.
It is interesting to notice that in the large $U$ limit 
the bandshift of the $(\sigma)$-band being
proportional to $B_{-\sigma}$	 is mainly determined by the hopping of the
$(-\sigma)$-quasiparticles.

The density, double-hopping and spinflip correlation functions reduce in this
limit to:
\begin{eqnarray} 
F_{-\sigma}^{(1)}
	&\stackrel{n<1,U\rightarrow\infty}{\llongrightarrow}& 
	-(1-\nmsi)\frac{\lb c\imsi^{\dagger} c\jmsi\rb^{2}}{1-n}, 
			\label{limes1}\\*
F_{-\sigma}^{(2)}
	 &\stackrel{n<1,U\rightarrow\infty}{\llongrightarrow}& 0, 
	 		\label{limes2}\\*
F_{-\sigma}^{(3)}
	 &\stackrel{n<1,U\rightarrow\infty}{\llongrightarrow}&
	 -\frac{\lb c\isi^{\dagger} c\jsi\rb
	\lb c\imsi^{\dagger} c\jmsi\rb}{1-n}.
			\label{limes3}
\end{eqnarray}

\textbf{(ii) Paramagnetism, large $U$ and less than half filling:}
In the case of paramagnetism we can combine (\ref{limes1})-(\ref{limes3}),
which yields for the bandwidth correction $F_{-\sigma}$:
\begin{equation}
\label{flu}
	F_{-\sigma}\quad
	\stackrel{n<1,U\rightarrow\infty}{\llongrightarrow}
	\quad\frac{4}{n(2-n)}
	\left[-\left(2-\frac{n}{2}\right)\frac
	{\lb c_{i\sigma}^{\dagger} c\jsi\rb^{2}}{1-n}\right].
\end{equation}
By introducing the spin-spin correlation function 
$\lb \vec{S}_{i}\vec{S}_{j}\rb $ this can be written as:
\begin{equation}
F_{-\sigma}\quad
	\stackrel{n<1,U\rightarrow\infty}{\llongrightarrow}
	\quad\frac{4}{n(2-n)}
	\left[\lb \vec{S}_{i}\vec{S}_{j}\rb -
	\frac{\lb c_{i\sigma}^{\dagger} c\jsi\rb^{2}}{2}\right].
\end{equation}
Here, the spin-spin correlation function is defined in the usual way:
\begin{equation}
\lb \vec{S}_{i}\vec{S}_{j}\rb=
	 \frac{1}{2}\lb S_{i}^{+}S_{j}^{-}\rb+
	    \frac{1}{2}\lb S_{i}^{-}S_{j}^{+}\rb
	   +\lb S_{i}^{z}S_{j}^{z}\rb,
\end{equation}
with:
\begin{equation}
	S_{i}^{+}=c_{i\uparrow}^{\dagger}c_{i\downarrow}, \,\,
	S_{i}^{-}=c_{i\downarrow}^{\dagger}c_{i\uparrow}, \,\,
	S_{i}^{z}=\frac{1}{2}(n_{i\uparrow}-n_{i\downarrow}).
\end{equation}

\textbf{(iii) Paramagnetism and half filling:}
For the exact half filled band ($n=1$) we can use (\ref{nicicjngleich1}) to
simplify the expressions for $B_{-\sigma}$ and $F_{-\sigma}$:
\begin{eqnarray}
	B_{-\sigma}&=&T_{0}\\
	F_{-\sigma}&=&\frac{12}{n(2-n)}\left[ 
	-\frac{\lb c\isi^{\dagger} c\jsi\rb^{2}}
	{8\lb n\isi n\imsi\rb-16\lb n\isi
	n\imsi\rb^{2}}\right].\label{f_half_filling}
\end{eqnarray}	
\textbf{(iv) Paramagnetism, half filling and large $U$:}
If we perform the large $U$ limit starting from the exact expression
(\ref{f_half_filling}) for $F_{-\sigma}$, a straightforward calculation yields:
%
\begin{equation}\label{f_hf_lu}
	F_{-\sigma}
	\stackrel{n=1,U\rightarrow\infty}{\llongrightarrow}
	-\frac{3}{Z_{1}}.
\end{equation}
At this point we want to emphasize that $n\rightarrow 1$ and the 
limit $U\rightarrow\infty$  do not commute. 
This can easily be seen with the following
relation for the hopping correlation function which is valid for $n\le 1$ and
$U\rightarrow\infty$:
\begin{equation} 
 	\lb c\isi^{\dagger} c\jsi\rb\le \frac{-1}{Z_{1}T_{1}}
	(1-n).
\end{equation}	
Therefore, expression (\ref{flu}) for $F_{-\sigma}$  vanishes as 
$n\rightarrow 1$ contrary to the correct result given by 
(\ref{f_hf_lu}).

\section{DISCUSSION AND RESULTS}\label{disc}
\begin{figure}
	\centerline{\psfig{figure=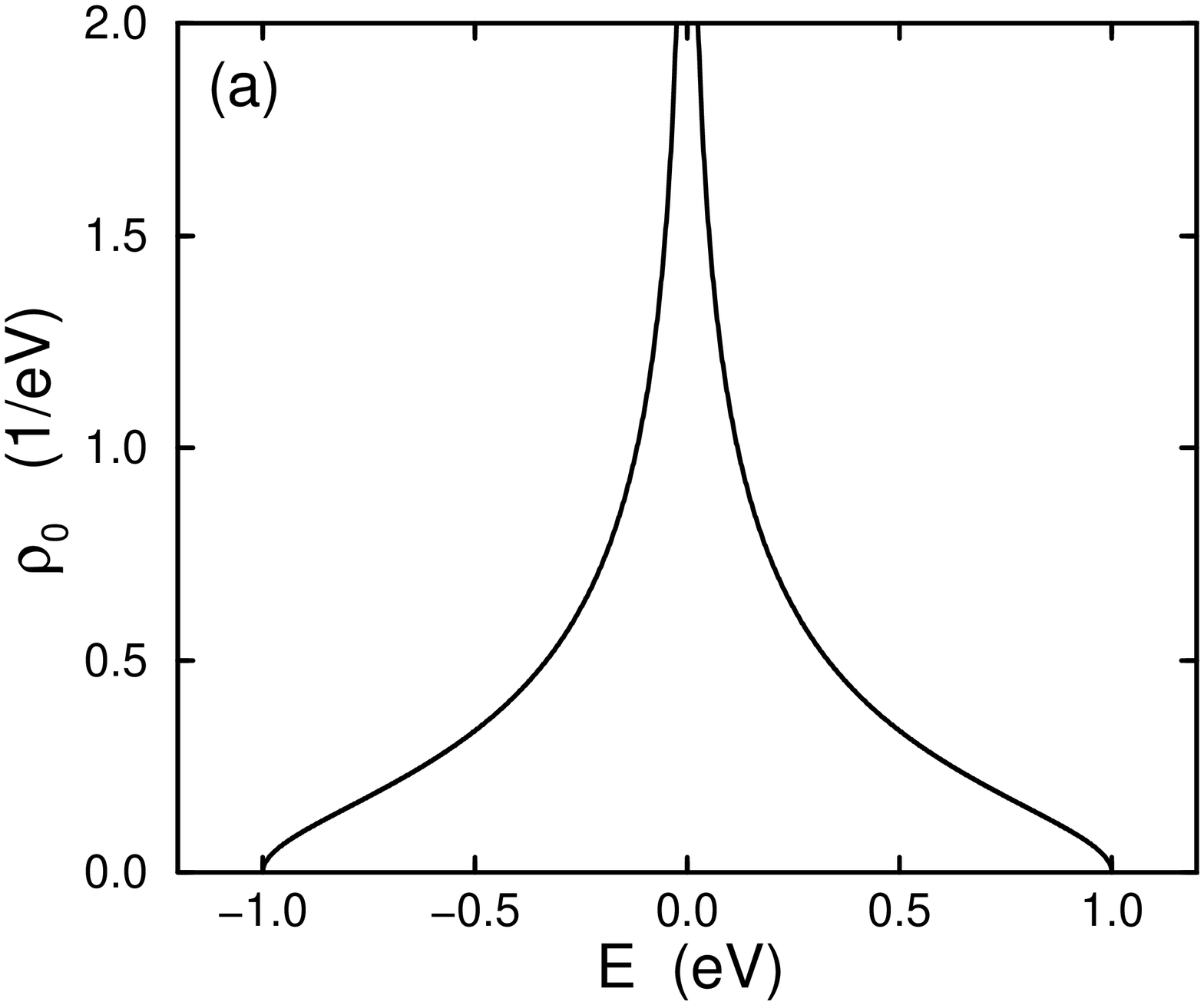,width=6cm,angle=270}
	\psfig{figure=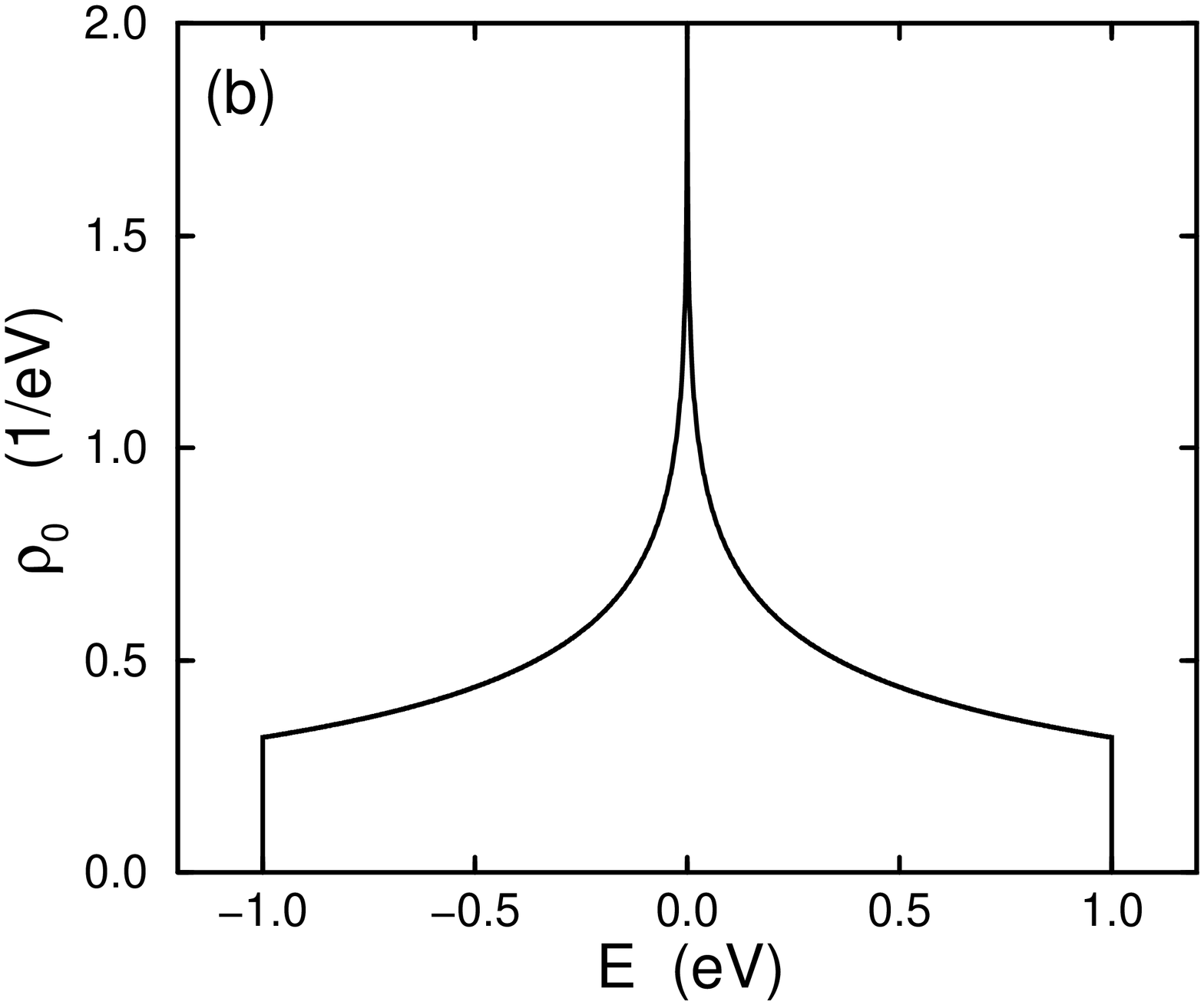,width=6cm,angle=270}}
\caption{Bloch density of states (BDOS) for (a) a three-dimensional 
bcc lattice 
and (b) a two-dimensional square lattice.\label{fig_bdos}}
\end{figure}

The calculations in three dimensions are done for a lattice with bcc
structure with the tight-binding dispersion given by:
\begin{equation}\label{ek3D}
	\ek= Z_{1}T_{1}\cos(k_{x}a)\cos(k_{y}a)\cos(k_{z}a) + T_{0}.
\end{equation}
\noindent a is the lattice constant, $Z_{1}$ the number of nearest neighbours and $T_{1}$
the hopping integral between neighbouring lattice sites. 
For the calculations in two dimensions we choose the dispersion of the
square lattice: 
\begin{equation}\label{ek2D}
\ek= \frac{Z_{1}T_{1}}{2}\bigg(\cos(k_{x}a)+\cos(k_{y}a)\bigg) 
	+ T_{0}.
\end{equation}
Throughout this section, we fix the Bloch-bandwidth to 
$W=-2Z_{1}T_{1}=2 \,\textrm{eV}$
and the center of gravity of the Bloch band  to $T_{0}=0 \,\textrm{eV}$.
The  corresponding BDOS are plotted in Fig.\,\ref{fig_bdos}.

Out of the infinite number of possible antiferromagnetic structures, 
we consider only the relatively simple and most commonly used ones, 
AFM-(110) and AFM-(AB),
which are shown in
Fig.\,\ref{fig_afm_conf}.
Within  the tight-binding approximation the hopping
integrals $T_{ij}^{\alpha\beta}$ read
and the corresponding Bloch energies (\ref{eps_afm})  are given by:
\begin{eqnarray}
	\epsilon_{AA}^{(110)}(\vec{k}) =\epsilon_{BB}^{(110)}(\vec{k})
	&=&\!\!\!T_{0}+
	\label{afm_ek} \\*
	& &\hspace{-21ex}2T_{1}\bigg[\cos\left(
	\!\frac{a}{2}(k_{x}\!+\!k_{y}\!-\!k_{z})\!\right)
	+\cos\left(\!\frac{a}{2}(-k_{x}\!+\!k_{y}\!+\!k_{z})
	\!\right)\!\bigg],
	\nonumber\\*[0.5cm]
      \epsilon_{AB}^{(110)}(\vec{k}) =
      (\epsilon_{BA}^{(110)})^{\ast}(\vec{k})   
     &=&\!\!\!2T_{1}e^{i\vec{k}\cdot(\vec{r}_{A}-\vec{r}_{B})}
	 \\*
	& &\hspace{-20ex}\times\bigg[\cos\left(\!\frac{a}{2}(k_{x}\!+\!k_{y}\!+\!k_{z})\!\right)
	+\cos\left(\!\frac{a}{2}(k_{x}\!-\!k_{y}\!+\!k_{z})
	\!\right)\!\bigg],\nonumber
\end{eqnarray}
and
\begin{eqnarray}
	\epsilon_{AA}^{(AB)}(\vec{k})&\equiv&0, \\*
	\epsilon_{AB}^{(AB)}(\vec{k})
	&=&T_{1}e^{i\vec{k}\cdot(\vec{r}_{A}-\vec{r}_{B})}
	\nonumber\\*
	& &\hspace{3ex}\times
	\cos(\frac{k_{x}a}{2})
	\cos(\frac{k_{y}a}{2})\cos(\frac{k_{z}a}{2}) .                 
\end{eqnarray}

\subsection{Paramagnetism}
\begin{figure}
	\centerline{\psfig{figure=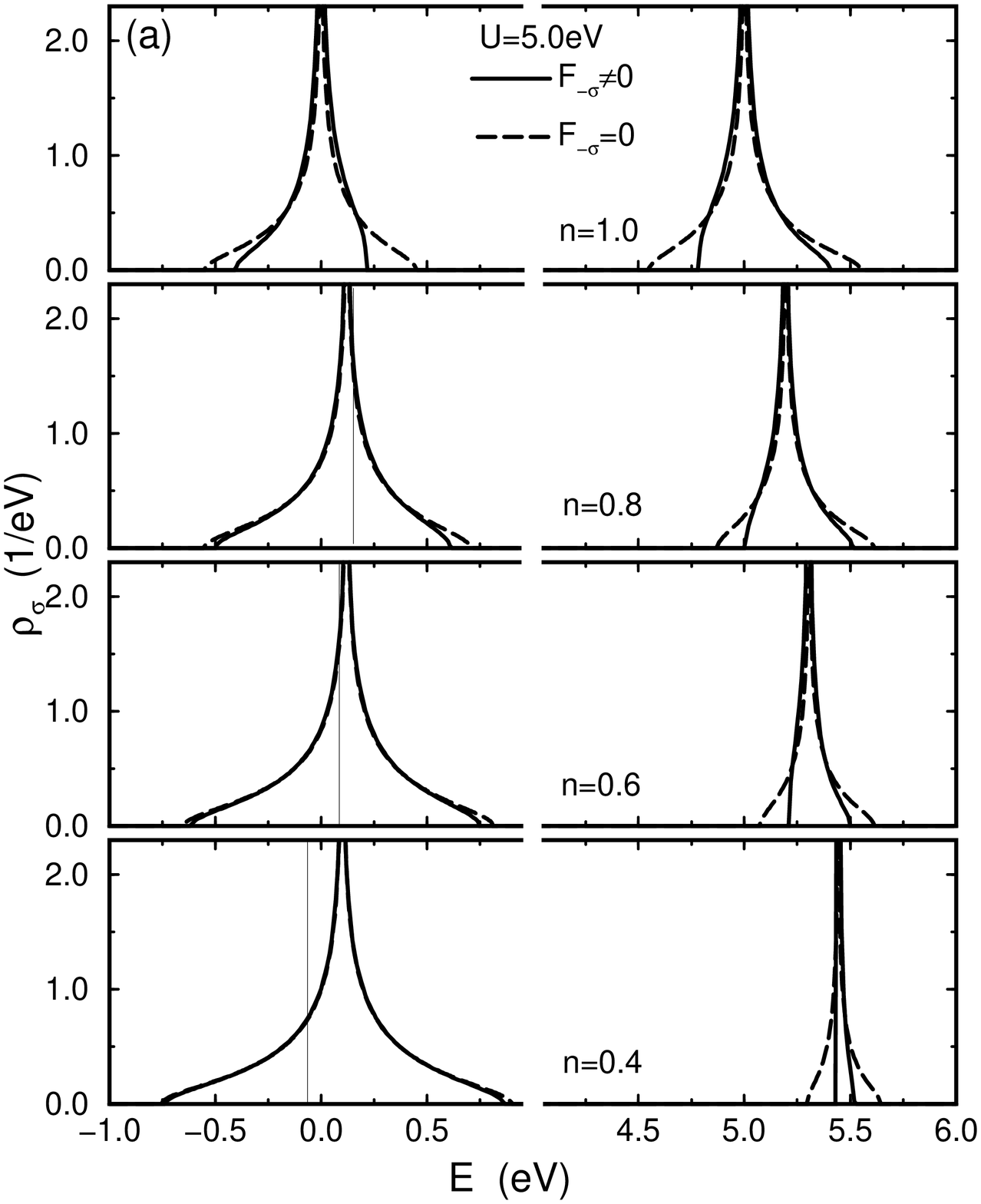,width=6cm}
	\psfig{figure=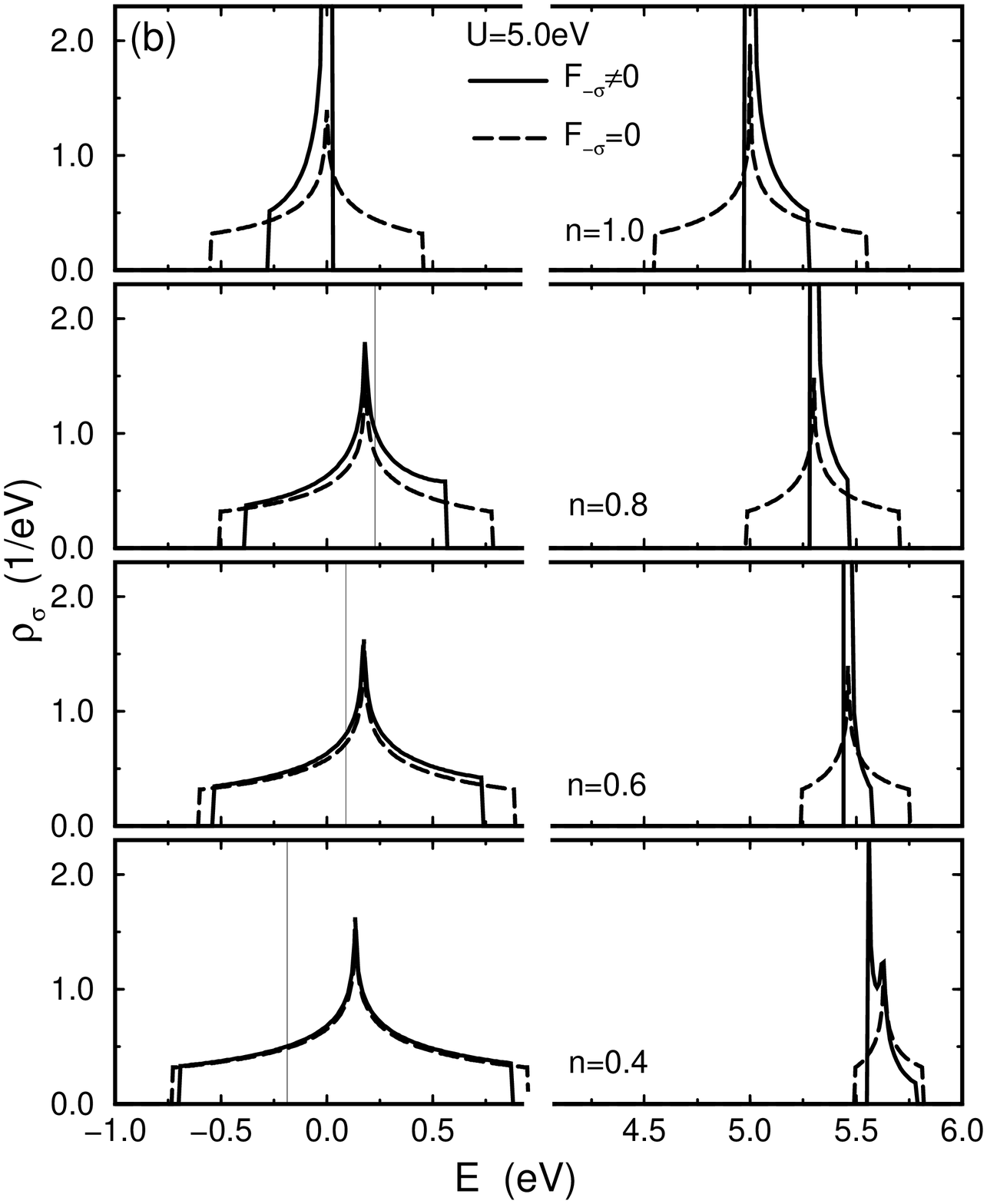,width=6cm}}
	\caption{Paramagnetic quasiparticle density of states (QDOS) 
	as a function of the 
	energy
	$E$ in (a) three and (b) two dimensions for different band occupations
	$n$ and the Coulomb interaction $U=5.0\,$eV. 
	Solutions with (solid lines) and 
	without (broken lines) the
	non-local part of the electronic self-energy are shown.
	The vertical lines indicate the positions of the
	chemical potential $\mu$. ($W=2.0\,$eV, $T=0\,$K)\label{fig_qdos_pm}}
\end{figure}

In Fig.\,\ref{fig_qdos_pm} the QDOS in three and two dimensions are plotted for
both cases, with and without the $\vec{k}$-dependent 
term $F_{\vec{k}-\sigma}$ of the self-energy.
The quasiparticle spectrum is divided into two parts, a low and a high energy
region. These two quasiparticle subbands (so-called Hubbard-bands)  are
separated by an energy amount comparable to the intraatomic Coulomb matrix 
element $U$.
The inclusion of the $\vec{k}$-dependent 
term $F_{\vec{k}-\sigma}$ into the 
self-energy leads to a band narrowing which gets stronger with increasing
band occupation. 
Since the centers of gravity of the subbands remain almost
constant the energy gap between the subbands increases. As can be seen in 
Fig.\,\ref{fig_qdos_pm}, the band narrowing effect gets  much stronger 
as the dimension is lowered from three to two. 
Since, in principle, magnetism is favoured by narrow energy bands
one expects the magnetic region to be enhanced by taking the non-locality of 
the self-energy into account.
The band narrowing effect holds for the attractive 
Hubbard model ($U<0$) as well. Contrary to the suggestions in \cite{SPR96},
the two
Hubbard bands do not melt together as the non-local part of the self-energy is
included.

\begin{figure}
	\centerline{\psfig{figure=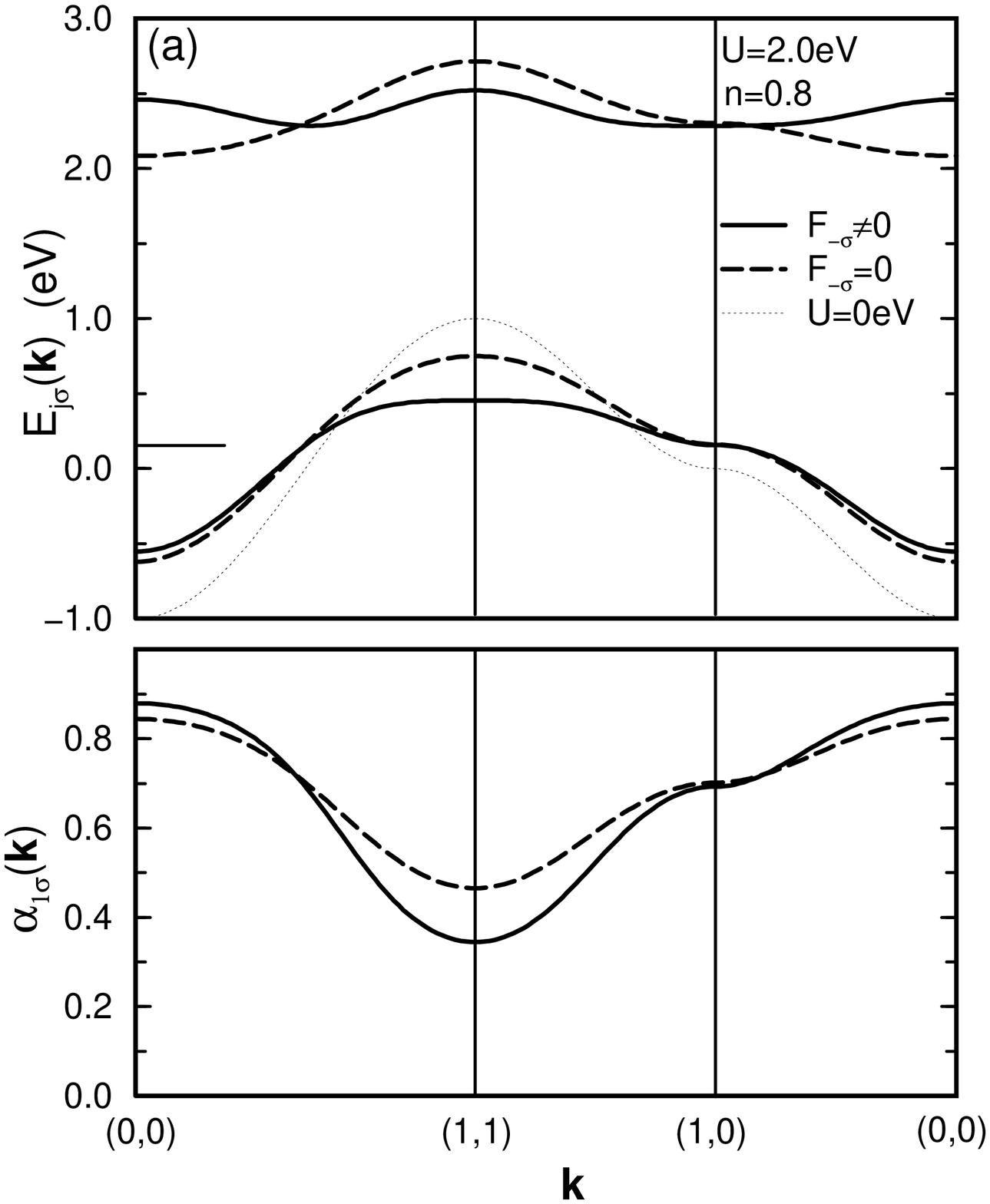,width=6cm}
	\psfig{figure=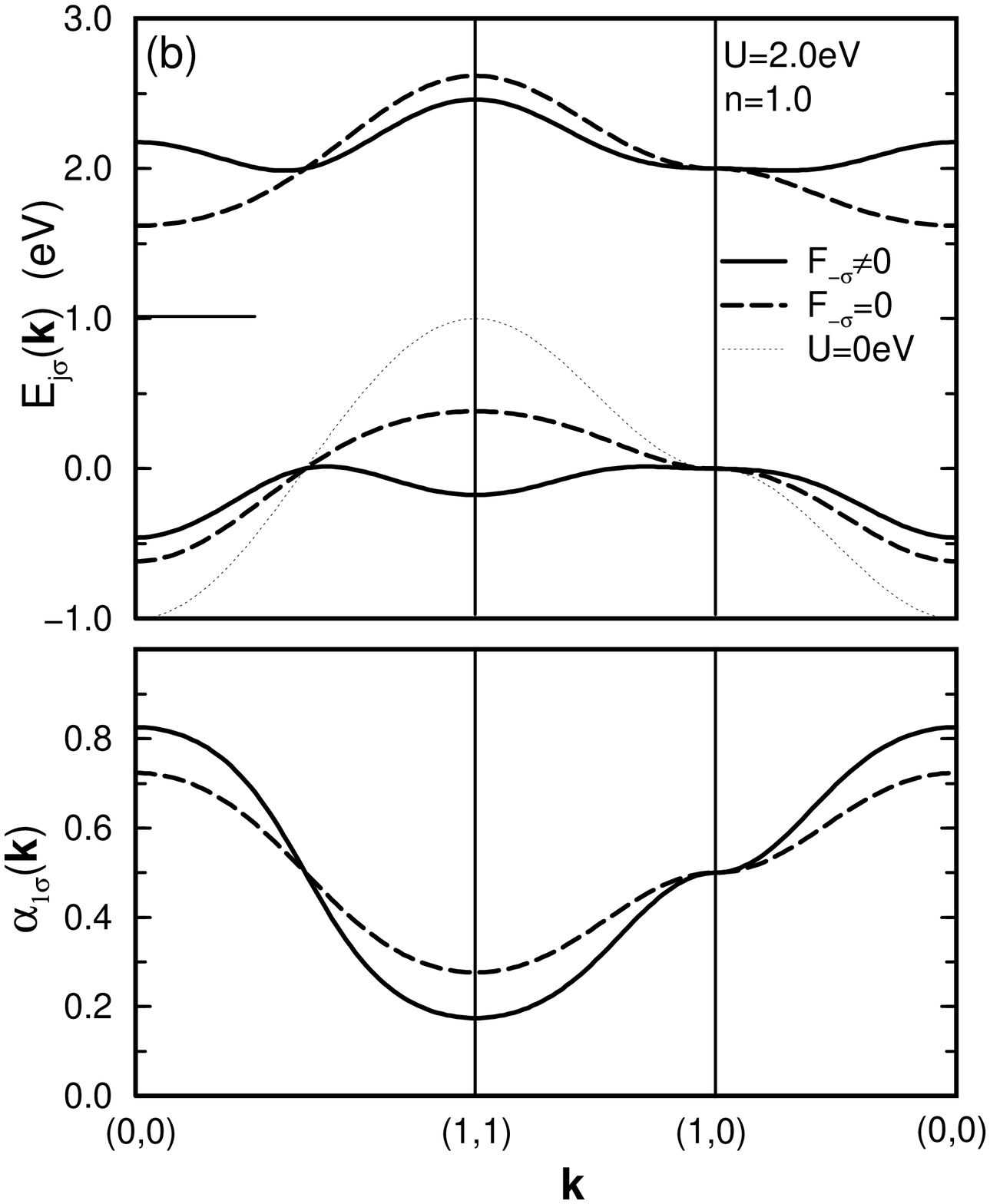,width=6cm}}
\caption{Paramagnetic quasiparticle dispersion 
	  $E_{j\sigma}(\vec{k})$ ($j=1,2$) 
	   and the spectral
         weight $\alpha_{1\sigma}(\vec{k})$ in two dimensions for  the
         band occupations (a) $n=0.8$ and  (b) $n=1.0$ and the Coulomb
         interaction $U=2.0\,$eV.
         Solid lines: Solutions with included non-local part of the self-energy.
         Broken lines: Only the local part of the self-energy is taken into
         account.
         The dotted curve corresponds to  
         the dispersion $\ek$ of the Bloch band. 
         The horizontal bars
         indicate the positions of the chemical potential $\mu$. The spectral
         weight of the upper quasiparticle subband is given by:
         $\alpha_{2\sigma}(\vec{k})=1-\alpha_{1\sigma}(\vec{k})$.
	   ($W=2.0\,$eV, $T=0\,$K)\label{fig_disp_pm}}
\end{figure}

Further insight into the effect of the non-local term of the self-energy  can
be obtained from the quasiparticle dispersion which is shown in 
Fig.\,\ref{fig_disp_pm}
for the two-dimensional system and moderate Coulomb interaction $U=W$.
The band narrowing of the lower Hubbard-band is mainly due to a flattening of
the dispersion near (1,1). The flattening extends to the saddle point (1,0) and
halfway to (0,0). For half filling  the inclusion of the $\vec{k}$-dependence
even leads to a local minimum of the dispersion at (1,1). 

Recently, this flattening of the bands has been  discussed in 
connection with
high temperature  superconductivity \cite{Dag94,BE95}, 
because similarly flat
dispersions are 
seen in angle resolved photoemission experiments on hole-doped 
cuprate superconductors. Beenen and Edwards \cite{BE95}
show that quasiparticle dispersions like the ones in Fig.\,\ref{fig_disp_pm}
compare rather well with quantum monte carlo simulations for a 
$4\times4$-cluster \cite{BSW94a,BSW94b}.  Similar results for one 
dimension are given in 
\cite{MEHM95}
where the authors argue that the inclusion of the non-local terms is 
crucial for a good agreement of the model calculation with   
exact diagonalization results for a 10-site Hubbard-ring.

For the complete information about the quasiparticle spectrum one has to look at
both, the quasiparticle energies $E_{j\sigma}(\vec{k})$  and the corresponding
spectral weights $\alpha_{j\sigma}(\vec{k})$.  The spectral weight 
$\alpha_{1\sigma}(\vec{k})$ of the lower
quasiparticle subband is shown in the lower part of Fig.\,\ref{fig_disp_pm}. The
spectral weight of the upper subband is given by 
$\alpha_{2\sigma}(\vec{k})=1-\alpha_{1\sigma}(\vec{k})$.  If one introduces
holes into the half filled system, there is a strong  transfer of spectral 
weight from the upper to  the lower  subband. This  weight transfer 
 is particularly prominent around  (1,1).

\begin{figure}
	\centerline{
	\psfig{figure=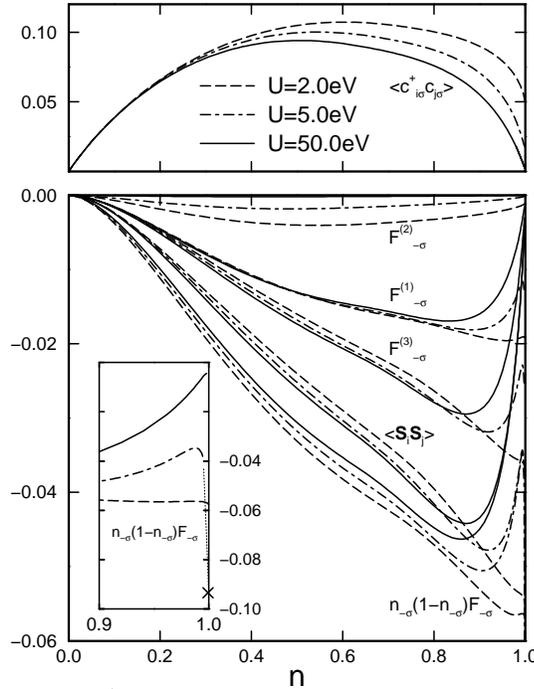,width=7cm}}
	\caption{
	Top: Hopping correlation function 
	$\lb c_{i\sigma}^{\dagger}c_{j\sigma}\rb$
	 between nearest neighbours in three
	dimensions as a
	function of the band occupation $n$ for various values of the Coulomb
	interaction $U$.
	Bottom: Density $F^{(1)}$, double hopping $F^{(2)}$, 
	spinflip correlation $F^{(3)}$ and the
	sum of these three correlation functions $\nmsi(1-\nmsi)F_{-\sigma}$ for
	the three-dimensional system as a function of $n$ for various $U$. 
	In addition, the spin-spin
	correlation function  $\lb \vec{S}_{i}\vec{S}_{j}\rb $ is shown.
	In the inset,  the limit $n\rightarrow 1$ of $\nmsi(1-\nmsi)F_{-\sigma}$
	is enlarged.($W=2.0\,$eV, $T=0\,$K)
	\label{fig_corr_pm}}
\end{figure}

Now, we want to focus on the bandwidth correction and the three correlation
functions contained therein. In Fig.\,\ref{fig_corr_pm} these correlation
functions are plotted for the three-dimensional case 
as a function of $n$ for three different values of $U$.
All correlation functions are negative and disappear, of course, in the empty
band limit ($n=0$) because of the lack of interaction partners. 

The double hopping correlation function is rather small and vanishes
completely for $U\rightarrow\infty$ due to the unfavourable double occupancies.
Since double hopping processes require a double occupied and an empty site, the
double hopping correlation function exhibits a weak maximum 
at quarter filling
and disappears at half filling and not too small values of $U$
(here and in the following we refer  to the absolute 
values of the correlation functions).
The double hopping correlation plays, therefore, a minor role in the following.

Of greater  importance with respect to the $\vec{k}$-dependence are the density
and the spinflip correlation. Both correlation functions increase in a wide
region of the band occupation $n$  monotonously with increasing $n$ and 
are, in this
region, nearly independent of $U$. Close to half filling and for $U>W$, however, 
this behaviour changes  and both correlation functions are strongly reduced.
This is because near half filling and not too small values of $U$ 
 the hopping between nearest neighbours
is suppressed. If the lower Hubbard-band gets filled, hopping is possible
only via double occupancies.

In the case of exact half filling ($n=1$), however, the density and spinflip
correlation, and therefore $F_{-\sigma}$,  have a finite value for all $U$. 
The behaviour of $n_{-\sigma}(1-n_{-\sigma})F_{-\sigma}$ is shown in the inset
in Fig.\,\ref{fig_corr_pm}. For $U=5.0\,$eV a smooth transition to the value at
half filling is indicated, although we do not find fully converged numerical
solutions as $n\arrowle1$. For $U\rightarrow\infty$ this transition becomes
discontinuous, indicating that the two limits $n\rightarrow 1$ and
$U\rightarrow\infty$ do  not commute (see Sec.\,\ref{sec_limits}).
For not too small $U$ and half filling, the numerical calculations  reproduce  
$F_{-\sigma}=-3/Z_{1}$ given by (\ref{f_hf_lu}) quite accurately.

From Fig.\,\ref{fig_corr_pm} one obtains that almost all the effect 
of the density and spinflip
correlation is gathered in the spin-spin correlation function 
$\lb\vec{S}_{i}\vec{S}_{j}\rb$ which indicates antiferromagnetic correlations
between nearest neighbours.

\begin{figure}
	\centerline{\psfig{figure=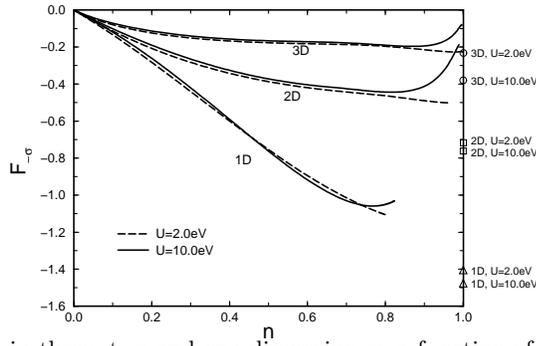,width=7cm,angle=270}}
	\caption{Bandwidth correction $F_{-\sigma}$ in three, 
	two and one dimension as a function of the band occupation $n$. 
	The values of $F_{-\sigma}$ for half filling $n=1$ are given
	on the right. ($W=2.0\,$eV, $T=0\,$K)
	\label{fig_fs_123d}}
\end{figure}

In Fig.\,\ref{fig_fs_123d} the bandwidth correction is shown in one, two and
three dimensions. The bandwidth correction clearly gets less important as the
dimension increases. This fits perfectly well into the picture developed by the
investigations of the Hubbard model for infinite dimensions where the
self-energy  becomes local \cite{MV89}. The
SDA reproduces the locality of the self-energy in the limit
$d\rightarrow\infty$.

\subsection{Spontaneous magnetism}
\begin{figure}
	\centerline{\psfig{figure=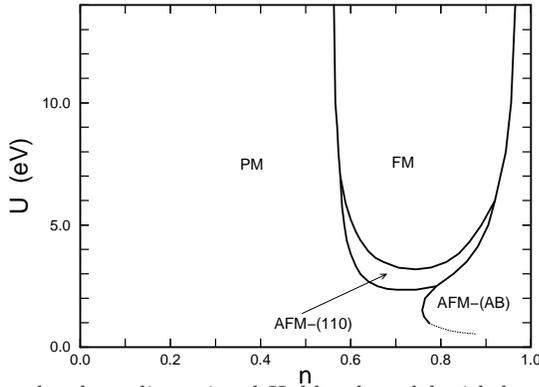,width=7cm,angle=270}}
	\caption{Magnetic phase diagram for the three-dimensional Hubbard-model
	with bcc lattice structure. $n$ is the band occupation and $U$ the
	intraatomic Coulomb matrix element. The bandwidth of the BDOS is fixed to 
	$W=2.0\,$eV.
	Four phases are considered:
	Paramagnetism (PM), ferromagnetism (FM) and antiferromagnetism in two
	configurations AFM-(110) and AFM-(AB). ($T=0\,$K)
	\label{fig_phase_3d}}
\end{figure}

\textbf{Magnetic phase diagram:} The global results of our calculations are
summarized in the magnetic phase diagram, which is shown in
Fig.\,\ref{fig_phase_3d}. We restrict the phase diagram to the region 
$0\le n \le 1$, since due to particle-hole symmetry the 
phase diagram is symmetric to the $n=1$ axis. 

Within the SDA, we find in wide parameter regions ferromagnetic and
antiferromagnetic solutions besides the ever existing paramagnetic one. 
If there is more than one mathematical solution of
the system, 
the physically stable one is indicated by the minimum free energy.
Spontaneous magnetism is possible only for $n$ greater than a critical band 
occupation $n_{c}(U)$. As $U$ increases $n_{c}$ decreases slightly without
getting smaller than $n_{0}=0.54$. 
This is consistent with  the results obtained by Kanamori \cite{Kan63}.
Here, we see the qualitative improvement of the SDA 
upon  a simple, but often used Hartree-Fock approximation (Stoner model) 
\cite{Sto36},  where  magnetic solutions can be found 
for all band occupations as long as the Coulomb interaction $U$ is 
strong enough \cite{NB89,PN95}.
The Hartree-Fock approximation is well known to overestimate the possibility of
spontaneous magnetism.

Besides a critical value for the band occupation, also a critical Coulomb
interaction $U_{c}$ is needed to build up spontaneous ferromagnetic solutions. 
$U_{c}^{\ix{FM}}$ has a minimum at $n=0.75$, where,  consistent with 
the Stoner criterion,  the chemical potential $\mu$ lies in the region 
of maximum density of states.
$U_{c}^{\ix{AFM-(110)}}$ is slightly bigger than the Bloch-bandwidth $W$. 
However, antiferromagnetism in the AFM-(110) configuration is energetically 
stable  only in a small
region below the ferromagnetic phase. Antiferromagnetism in the 
AFM-(AB) configuration is observed for band occupations close to
half-filling. 
We find stable AFM-(AB) solutions down to very small values of $U$. However,
the question, if AFM-(AB) solutions exist even for $U\rightarrow 0^{+}$ 
(as for the sc lattice, see for example \cite{SSST89}) can not 
reasonably be addressed within our strong coupling approach.
In the limit of large $U$,
the region, where antiferromagnetism in the AFM-(AB) configuration is stable
against ferromagnetism, is reduced to a small corridor around $n=1$.
Altogether, antiferromagnetism is
favoured by moderate Coulomb interaction $U\approx W$. 
 

The stability of the AFM-(AB) phase for the  half filled band $n=1$
can be explained as
follows: Via second order perturbation theory the Hubbard model at half filling
can be
transformed into an effective Heisenberg model with
exchange integrals equal to 
$J_{ij}=-2(T_{ij})^{2}/U \,\,\,(i\neq j)$ \cite{And63}. 
Therefore, the total
energy of the system is lowered by virtual hopping processes 
where an electron hops from lattice site $i$ to a neighbouring site $j$ and
back to $i$ again \cite{PN95}. For this virtual processes to be possible, the
quasiparticles on neighbouring sites need to have reversed spins 
because of the Pauli exclusion principle. Thus the lowering 
of the total energy  is  determined by the number of neighbouring 
sites occupied with reversed spin quasiparticles. This number is highest 
in the AFM-(AB) configuration, followed
by AFM-(110), PM and being zero for FM. This corresponds exactly 
to the results of our numerical calculations. 

Applying a modified perturbation theory (SOPT-HF: second order 
perturbation theory around the Hartree-Fock solution) 
Bulk and Jelitto \cite{BJ90} derive a
qualitatively very similar phase diagram. 
In particular, their calculations yield
similar values for the critical parameter $n_{c}$ and $U_{c}^{\ix{FM}}$. This
is a remarkable result, because perturbation theory is usually restricted to
very small values of $U$, whereas the SDA does best in the strong coupling
region. 
The qualitative agreement between SDA and modified
perturbation theory, therefore, leads to the conclusion that both theories
produce reliable results with respect to magnetism 
for moderate Coulomb interaction $U\simeq W$ as well. 
\\

\begin{figure}
	\centerline{\psfig{figure=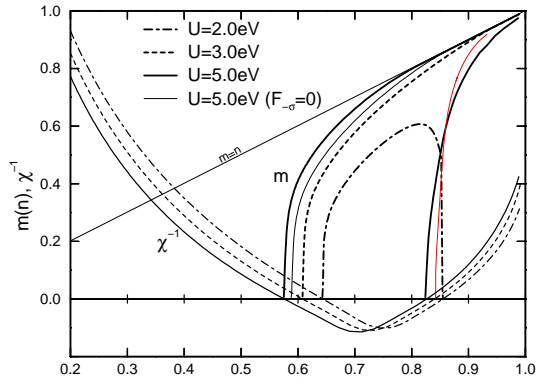,width=7cm,angle=270}}
	\caption{Magnetization $m$ as a function of the band occupation $n$  for
	different Coulomb interactions $U$. In addition, the inverse static
	paramagnetic susceptibility $(\chi^{(0)}(n,T))^{-1}$ 
	is shown in arbitrary units.($W=2.0\,$eV, $T=0\,$K)
	\label{fig_m_n_U_fm}}
\end{figure}

\textbf{Ferromagnetism:}  In Fig.\,\ref{fig_m_n_U_fm} the magnetization $m$
of the ferromagnetic system is
plotted as a function of $n$ for various values of $U$. In addition the inverse
static, paramagnetic susceptibility $(\chi^{(0)}(n,T))^{-1}$ is shown. 
$(\chi^{(0)}(n,T))^{-1}$
has either none or two roots which correspond to the instabilities of the system
against ferromagnetic order. The roots of $(\chi^{(0)}(n,T))^{-1}$, which has
been calculated from the paramagnetic solution only, indicate exactly the second
order phase transitions between paramagnetism and ferromagnetism.
As $U$ increases the magnetic solutions start from lower  band occupations and
show a higher magnetization. However, there are two different behaviours: For
$U<2.8\,$eV  the magnetization curves have a maximum around $n\simeq 0.8$ and
exist only between the two roots of  $(\chi^{(0)}(n,T))^{-1}$. 
But this behaviour only occurs in a parameter region,
where ferromagnetism is unstable 
against paramagnetism. Above $U\ge 2.8\,$eV we find two ferromagnetic
solutions FM1 and FM2, which start at the roots of $(\chi^{(0)}(n,T))^{-1}$ 
and exist both until $n=1$. By looking at the free energy we can conclude that
FM1 is always energetically favourable against FM2 \cite{GN88}. 
As $n$ increases, the FM1
solutions reach the ferromagnetic saturation $m=n$. In the limit 
$U\rightarrow\infty$ the system is fully polarized above $n=0.65$.

In Fig.\,\ref{fig_m_n_U_fm} we also show a magnetization curve  for $U=5\,$eV
without the $\vec{k}$-dependent term $F_{\vec{k}-\sigma}$ of the self-energy.
As expected, magnetism is favoured by inclusion of the $\vec{k}$-dependence.
Nevertheless, in three dimensions the influence of the $\vec{k}$-dependence
on the magnetic properties is rather small. The qualitative
behaviour of the system with and without $F_{\vec{k}-\sigma}$ is very similar.
The same is valid in the case of antiferromagnetism.

\begin{figure}
	\centerline{
	\psfig{figure=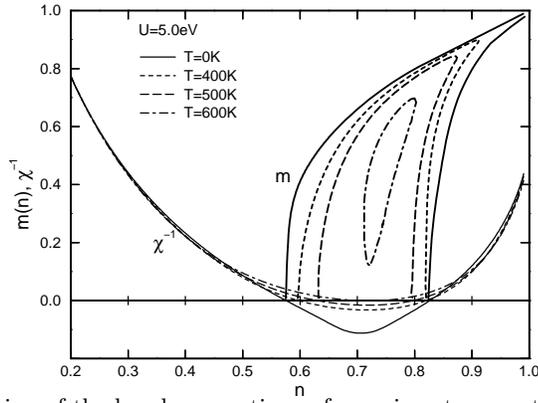,width=7cm,angle=270}}
	\caption{Magnetization $m$ as a function of the band occupation $n$ for
	various temperatures $T$. In addition, the inverse static
	paramagnetic susceptibility $(\chi^{(0)}(n,T))^{-1}$ 
	is shown in arbitrary units.
	($U=5.0\,$eV, $W=2.0\,$eV)\label{fig_m_n_T_fm}}
\end{figure}

\begin{figure}
	\centerline{
	\psfig{figure=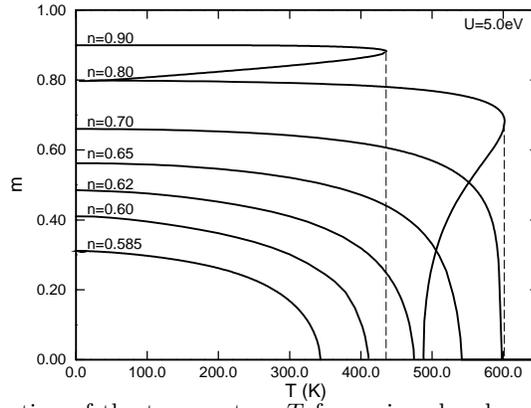,width=7cm,angle=270}}
	\caption{Magnetization $m$ as a function of the temperature $T$ for
	various band occupations $n$. The dashed vertical lines indicate first
	order phase transitions. ($U=5.0\,$eV, $W=2.0\,$eV)
	\label{fig_m_T_n_fm}}
\end{figure}

We now want to focus on the temperature dependence of the magnetization which is
plotted in Fig.\,\ref{fig_m_n_T_fm} as a function of $n$ for different
temperatures $T$ and in Fig.\,\ref{fig_m_T_n_fm} as a function of $T$ and
various band occupations $n$.
As $T$ increases the magnetization clearly decreases.  Near half filling the
two ferromagnetic solutions FM1 and FM2 melt together. This causes first order
phase transitions of the magnetization as a function of $T$ for band 
occupations close to
$n=1$ (see for example the $n=0.9$ curve in Fig.\,\ref{fig_m_T_n_fm}). Note
that, again, the energetically stable solution is always 
the one with the higher magnetic moment. 
In Fig.\,\ref{fig_m_T_n_fm} one can see that there are magnetic
solutions  that exist for higher temperatures than all Curie-temperatures
corresponding to  second order phase transitions. This explains the island-like
behaviour of the $T=600$K curve in Fig.\,\ref{fig_m_n_T_fm}.
For $n \le 0.7$ all magnetization curves $m(T)$ show a second order phase
transition with a shape very similar to a Brillouin function, as in localized
moment Heisenberg ferromagnets. 

\begin{figure}
	\centerline{\psfig{figure=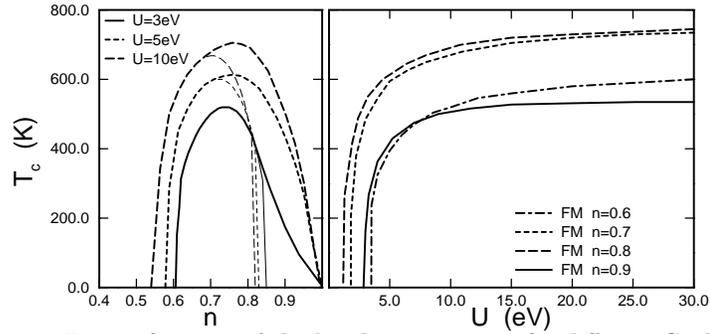,width=9cm,angle=270}}
	\caption{(a) Curie-temperature $T_{c}$   as a function of the band
	occupation $n$ for different Coulomb interactions $U$. The thin dashed
	lines correspond to the roots of the inverse static 
	susceptibility $(\chi^{(0)}(n,T))^{-1}$.
	(b) Curie-temperature $T_{c}$   as a function of $U$ for various $n$.
	($W=2.0\,$eV)
	\label{fig_Tc_fm}}
\end{figure}

From the magnetization curves $m(T)$ as well as from $(\chi^{(0)}(n,T))^{-1}$
we can determine the Curie temperatures $T_{C}$ which are shown in
Fig.\,\ref{fig_Tc_fm}. As a function of $n$ the Curie-temperatures are highest
in the region around $n\simeq 0.75$. Because the two ferromagnetic solutions
FM1 and FM2 melt together for very  low temperatures as $n$ approaches 
half filling, the Curie-temperatures reduce to zero at $n=1$. In 
Fig.\,\ref{fig_Tc_fm}(a) we also show the roots of the inverse static 
susceptibility. The system exhibits first order phase transitions as soon as
both curves are disjunct. 

For a given band filling $n$, the Curie temperature increases with the Coulomb
interaction $U$, but saturates  in the limit
$U\rightarrow\infty$ at a finite value which depends on $n$. We want to
emphasize that this behaviour is a decisive improvement upon the Stoner model
of band magnetism, for which $T_{C}$ is unrealistically high and 
increases linearly with $U$ \cite{NB89}.

\begin{figure}
	\centerline{\psfig{figure=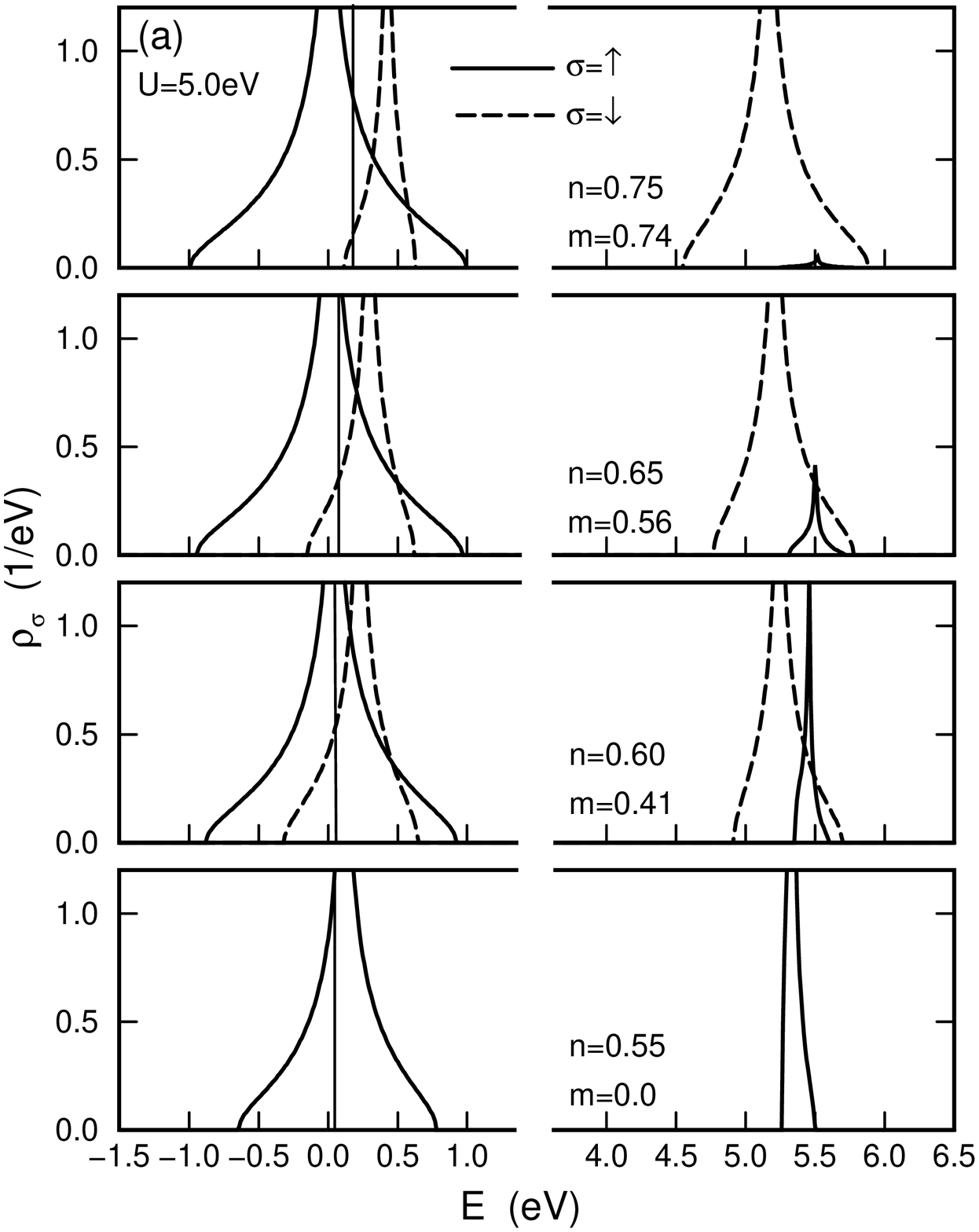,height=8cm,angle=0}
	\psfig{figure=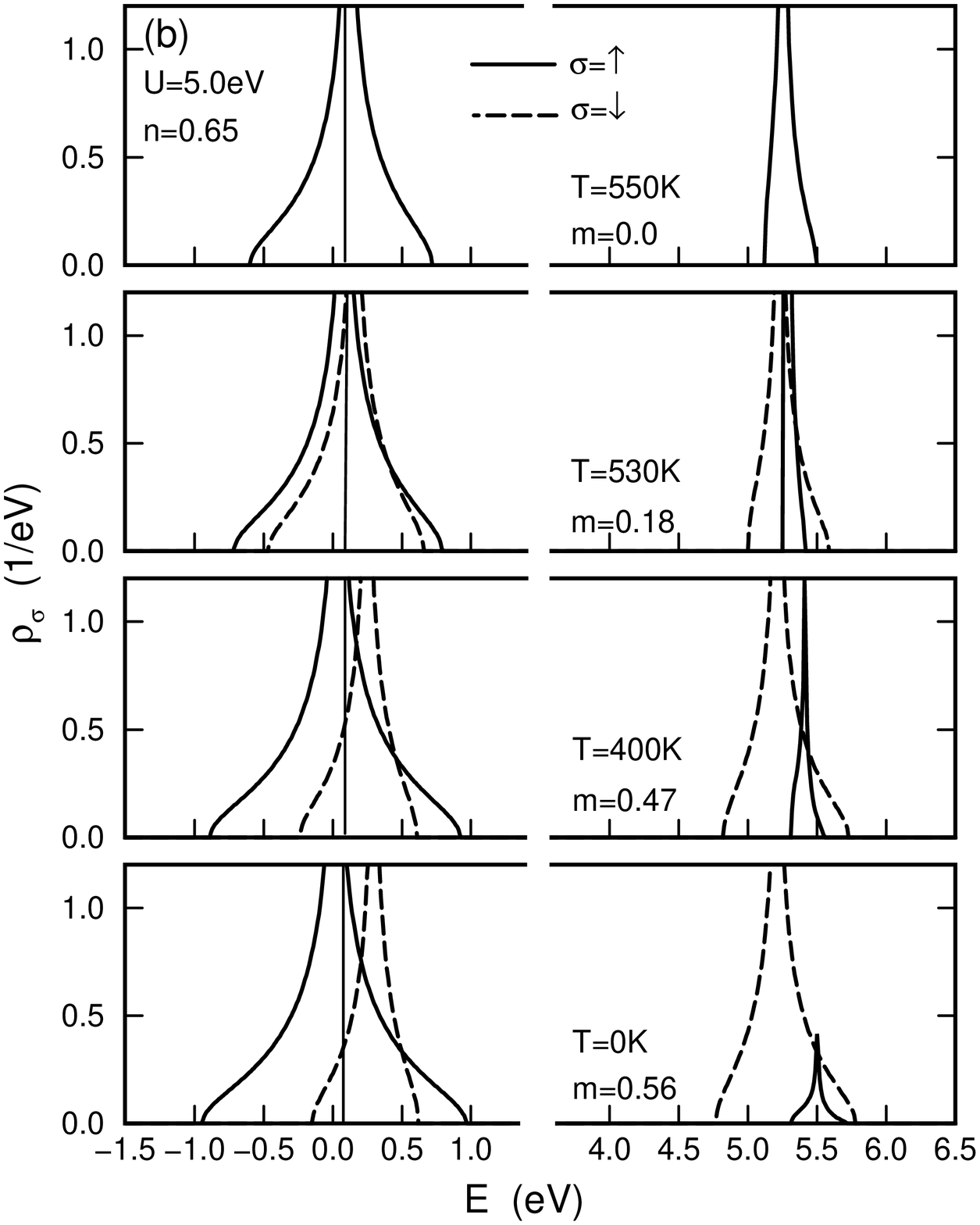,height=8cm,angle=0}
	\psfig{figure=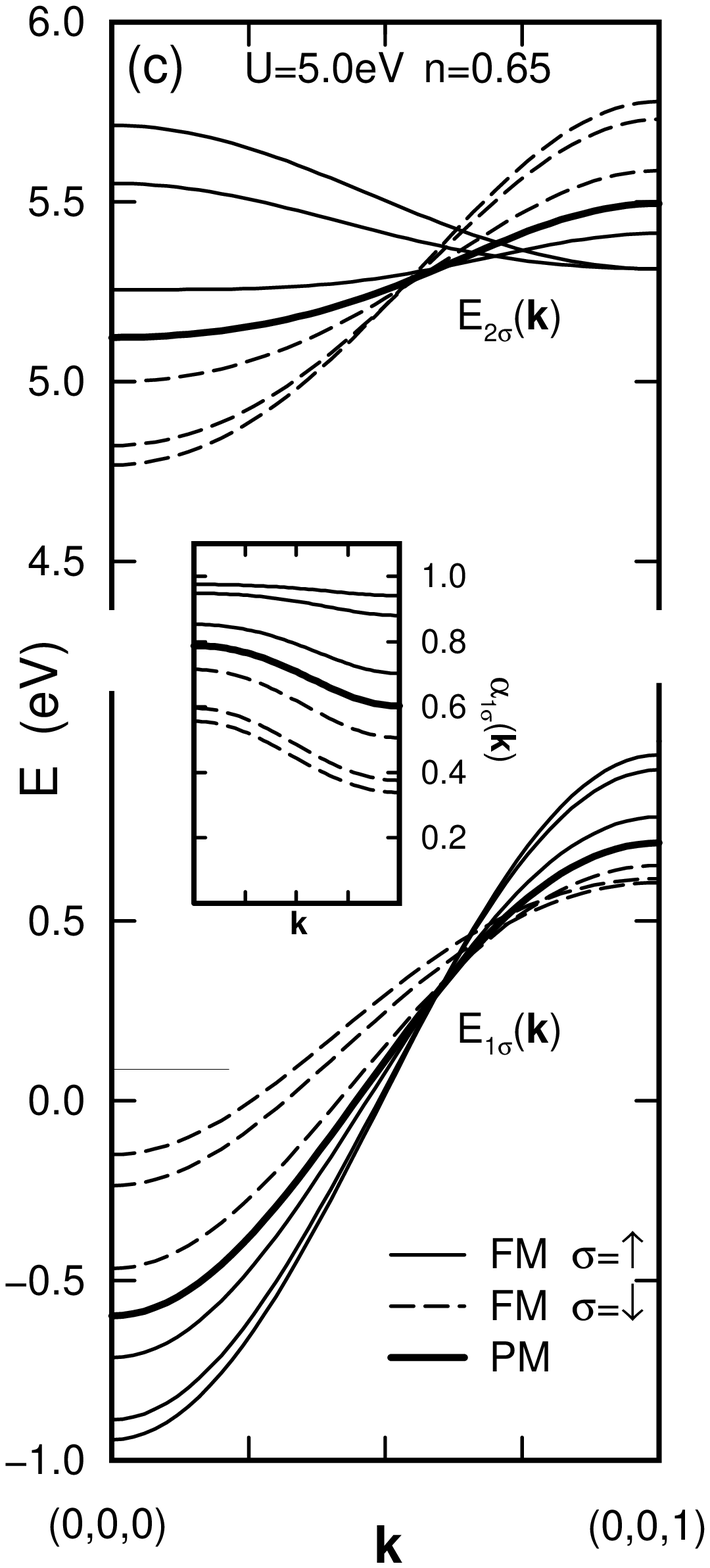,height=8cm,angle=0}}
	\caption{(a) QDOS $\rho_{\sigma}(E)$ as a 
	function of the energy $E$ for four
	different band fillings $n$. Solid lines correspond to the majority spin
	($\sigma=\uparrow$), broken lines to the minority spin direction
	($\sigma=\downarrow$).
	The vertical lines indicate the positions of
	the chemical potential $\mu$.
	(b) QDOS $\rho_{\sigma}(E)$ for various temperatures $T$. 
	(c) Quasiparticle dispersion $E\jsi(\vec{k})$ for the same parameter as
	in (b) along the (0,0,1)-direction of the first Brillouin zone. 
	In the inset, the spectral weight $\alpha_{1\sigma}(\vec{k})$ is shown.
	From the outside to the inside: $T=0\,$K, $T=400\,$K,
	$T=530\,$K und $T=550\,$K; the thick
	line corresponds to the paramagnetic  dispersion. 
	($W=2.0\,$eV, $T=0\,$K)\label{fig_qdos_fm}}
\end{figure}

For a better understanding of the macroscopic magnetic properties 
like $m$ and $T_{C}$, it is useful to look at the 
quasiparticle density of states and quasiparticle dispersion.  These are
plotted in Fig.\,\ref{fig_qdos_fm} for $T=0\,$K and $U=5.0\,$eV.
In Fig.\,\ref{fig_qdos_fm}(a) the QDOS is shown for four different band
occupations. For $n=0.55$ there is no ferromagnetic solution and the QDOS
consists of two quasiparticle peaks corresponding to the lower and upper
Hubbard-bands. If the band occupation exceeds $n=0.58$, however, an additional spin-splitting 
occurs between $\rho_{\uparrow}$ and $\rho_{\downarrow}$ which leads to
ferromagnetism. There are essentially two distinct correlation effects in the
QDOS:  
An energy shift between the centers of gravity of the $\uparrow$ and the
$\downarrow$ subbands (Stoner shift) and a deformation of the density of 
states. 
The band deformation  is directly connected to a transfer of spectral
weight between the upper and lower quasiparticle subbands.
The energy shift and the deformation of the  band are strongly $n$ and 
$T$ dependent.
Both effects together determine the spin-asymmetry of the QDOS and,
therefore, the magnetic properties of the system. In particular, near the upper
edge of the lower quasiparticle subband, the interplay between the band shift
and the band deformation leads to an inverse exchange
splitting, i.e the $\downarrow$ quasiparticle states are energetically below
the $\uparrow$ states. This is because the band narrowing of the lower minority
band overcompensates the Stoner shift. 

Qualitatively, the band deformation and the transfer of spectral weight 
is easy to understand: 
The lower quasiparticle
band results from electron hopping over lattice sites which are not occupied by
electrons with opposite spin. For non-zero
magnetization, the number of these sites becomes spin dependent. 
The probability of finding a majority electron on an otherwise empty site is
enhanced for $m>0$. Thus the lower $\rho_{\downarrow}$-band is narrowed as $m$
increases and finally vanishes  in the ferromagnetic saturation and
$n\rightarrow1$.
The lower majority subband, however,  is broadened with decreasing
number of minority particles until it
becomes, for $m=n$, equal to the BDOS because of missing interaction partners.
As can be seen analytically in the strong coupling limit, 
the weight of the lower $(\sigma)$-spin quasiparticle subband roughly scales 
with $(1-n_{-\sigma})$. The weight of the upper subband, where the
hopping of the  electrons is essentially via lattice sites which are already 
occupied by a reversed spin electron, scales with $n_{-\sigma}$.

The temperature  dependence of the QDOS and the corresponding quasiparticle
dispersions are shown in Fig.\,\ref{fig_qdos_fm}(b),(c). The spin asymmetry of
the partially filled lower quasiparticle subband decreases with increasing
temperature. This is, again, partly due to a Stoner-shift and partly due to a
deformation of the subbands. For temperatures above $T_{C}$, the
spin-splitting, of course, vanishes, but the upper and lower parts of the
spectrum remain separated by a gap of order $U$. 

The spin-splitting of the quasiparticle dispersion (Fig.\,\ref{fig_qdos_fm}(c))
scales with the magnetization. 
However, the actual amount of the spin-splitting strongly
depends on the position in the Brillouin zone. There even is  a region in
$\vec{k}$-space, where the quasiparticle energies are nearly independent of
$T$. Near the upper edge of the lower quasiparticle subband the above mentioned
inverse exchange
splitting is clearly visible.

In the inset in Fig.\,\ref{fig_qdos_fm}(c) the spectral weight of the lower peak
in the spectral density is shown. For $T=0\,$K the spectral weight is strongly
spin dependent. As the temperature increases, a redistribution of spectral
weight between the upper and lower subband leads to a decrease of this
asymmetry.  From Fig.\,\ref{fig_qdos_fm}(c) follows that a conventional
$E_{\sigma}(\vec{k})$ band-structure representation does not contain the full
information about the underlying system, because details about the  
distribution of spectral weight are missing. 

\begin{figure}
	\centerline{\psfig{figure=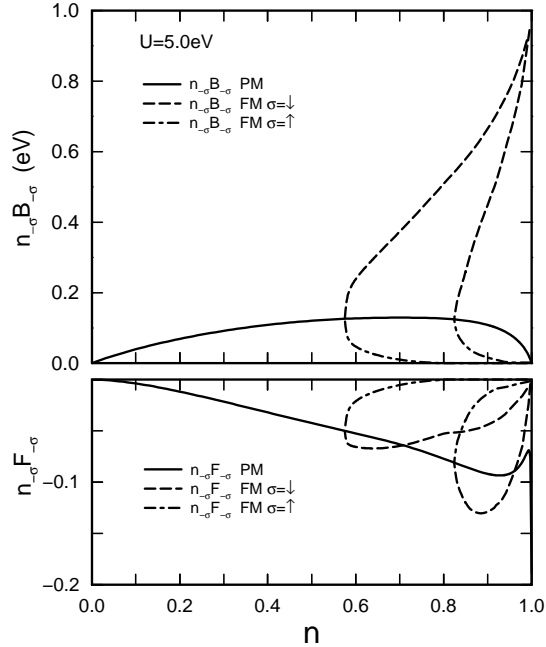,width=7cm}}
	\caption{Effective bandshift $\nmsi B_{-\sigma}$ and effective bandwidth
	correction $\nmsi F_{-\sigma}$ as a function of the band occupation $n$
	for the paramagnetic (PM) and the ferromagnetic (FM) phase.
	($W=2.0\,$eV, $T=0\,$K)}
	\label{fig_bf_fm}
\end{figure}

The spin asymmetry of the QDOS is due to the correction term
$B_{\vec{k}-\sigma}$ in the self-energy which is given by the bandshift 
$B_{-\sigma}$ and the bandwidth correction $F_{-\sigma}$. With respect to the
strong correlation limit, we have plotted in Fig.\,\ref{fig_bf_fm} the effective
bandshift $n_{-\sigma}B_{-\sigma}$ and the effective bandwidth correction 
$n_{-\sigma}F_{-\sigma}$ as a function of $n$.
Note that $n_{\downarrow}B_{\downarrow}$ is the effective bandshift of the
lower $\uparrow$-subband. The same holds for $n_{-\sigma}F_{-\sigma}$.

For $n$ larger than $n=0.58$ the effective bandshift becomes strongly spin
dependent, allowing, therefore, for the existence of ferromagnetism. 
The effective bandshift of the majority band  becomes very small and disappears
in the ferromagnetic saturation due to $n_{\downarrow}=0$.
The effective bandshift of the minority band, however, increases rapidly and
tends towards $W/2=1.0\,$eV as $n\rightarrow1$. The minority band  gets,
therefore, shifted above the Fermi edge.

The spin asymmetry of the effective bandwidth correction is less pronounced.
This is because only the density correlation is spin dependent. Further, 
$n_{-\sigma}F_{-\sigma}$ vanishes in the case of half filling. This is
consistent with the analytical results of section~\ref{sec_limits}.
\\

\begin{figure}
	\centerline{\psfig{figure=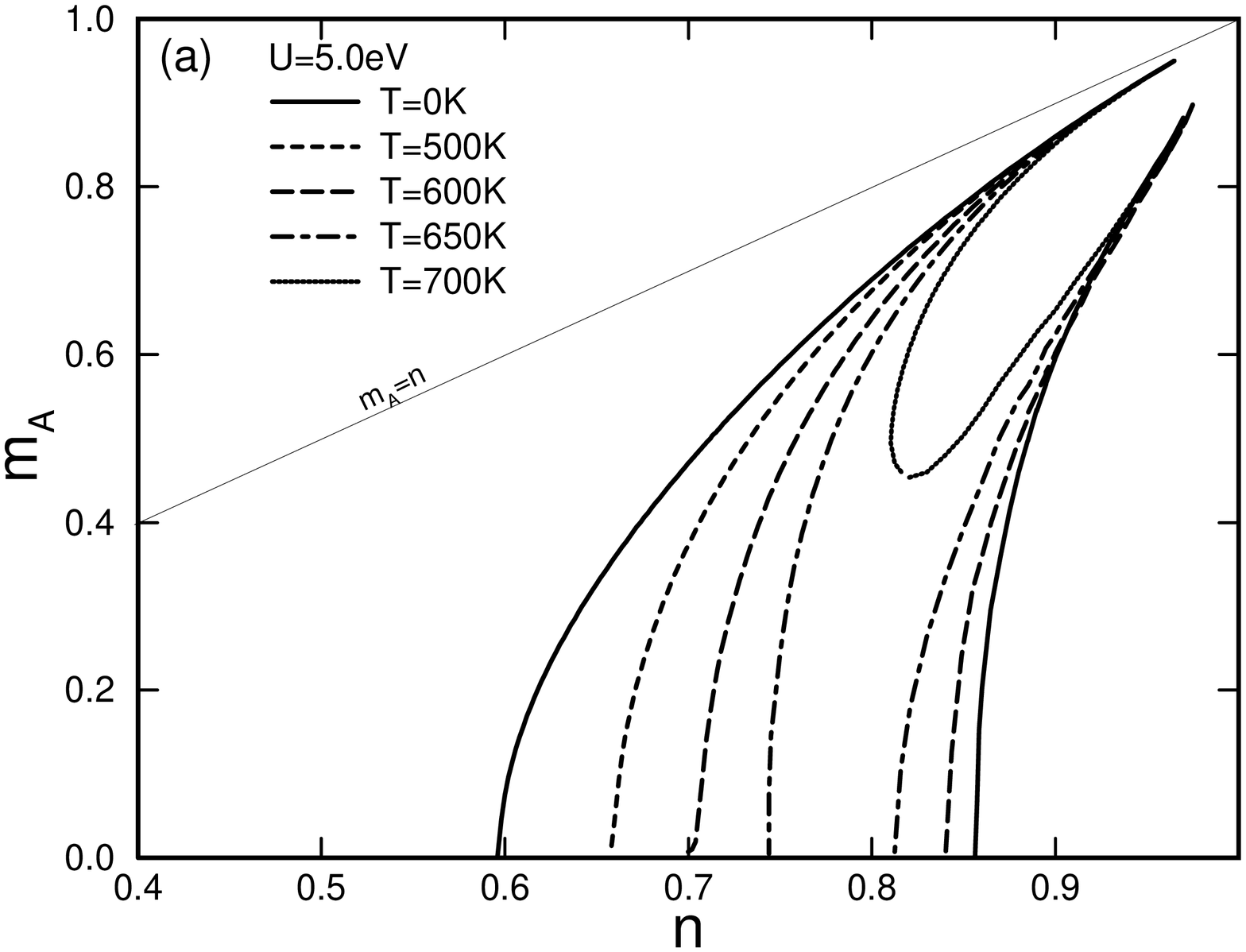,width=7cm,angle=270}
	\psfig{figure=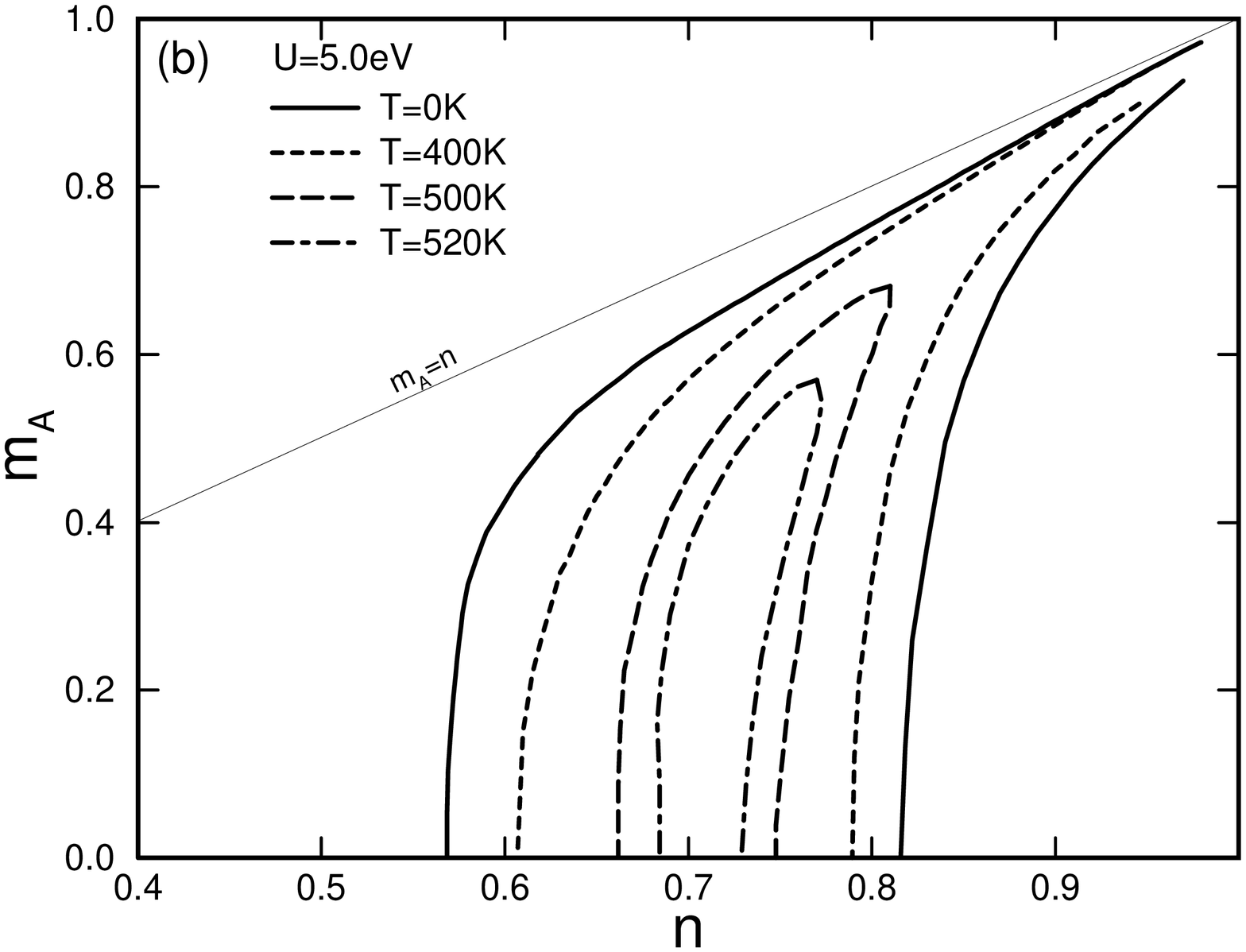,width=7cm,angle=270}}
	\caption{Sublattice magnetization $m_{A}$ as a function of the band
	occupation $n$ for various temperatures $T$; (a) AFM-(AB), (b) AFM-(110).
	($U=5.0\,$eV, $W=2.0\,$eV)
	\label{fig_m_n_T_afm}}
\end{figure}

\textbf{Antiferromagnetism:}
The sublattice magnetizations $m_{A}$ for different temperatures $T$ 
are plotted in 
Fig.\,\ref{fig_m_n_T_afm} (a) and (b) 
for the AFM-(110) and the AFM-(AB) configurations
respectively.  
The most striking difference to the ferromagnetic system is that the sublattice
magnetizations  never reach the
saturation $m_{A}=n$. This is because in an antiferromagnetic system the spin
up and spin down density of states occupy exactly the same energy region.
Filling the subbands with particles up to the Fermi level, therefore, 
always leads to a non-vanishing 
number of particles with minority spin, resulting in $m_{A}<n$. 

The temperature  dependence of $m_{A}$ in the AFM-(110) phase is quite
similar to the one in the case of ferromagnetism. 
This does not hold for the AFM-(AB) configuration, which is particularly stable
near half filling. The two solutions AFM1 and AFM2  in the AFM-(AB) phase stay
separated and do not melt together until fairly high temperatures. 
This can be understood  by means of the sublattice
density of states which are discussed later. The chemical potentials $\mu_{1}$,
$\mu_{2}$ of the  AFM1 and AFM2 solutions are separated in the AFM-(AB) phase by
the so-called Slater gap. For AFM1 and AFM2 to melt together, 
the broadening of the Fermi function has to be comparable to the size of the 
Slater gap.

In Fig.\,\ref{fig_Tn} the N\'{e}el temperatures  
are shown  as a function of Coulomb interaction $U$. In the AFM-(AB)
configuration the N\'{e}el temperatures are highest for moderate Coulomb
interaction, they decrease with increasing $U$ in the strong coupling region and  
converge to  a finite value as $U\rightarrow\infty$.
The maximum is most pronounced for band occupations close to half filling. 
As already mentioned, for the half filled band the Hubbard model 
can be transformed via
second order perturbation theory  into an effective Heisenberg model with
antiferromagnetic exchange integrals 
$J_{ij}=-2(T_{ij})^{2}/U$  \cite{And63,CSO77}. 
Thus, using the  mean field approximation for the
Heisenberg model, 
one finds $T_{N}\sim 1/U$. 
Close to half-filling  and for not too small values of $U$  the 
$T_{N}(U)$ curves in the AFM-(AB) phase show, at least qualitatively,
this behaviour.
In the AFM-(110) configuration, which
is built by  alternating ferromagnetic planes, the decrease of the N\'{e}el 
temperatures
as the Coulomb interaction increases  is only weak. 
Here again, antiferromagnetism 
in the AFM-(110) configuration  behaves quite similar to ferromagnetism.

\begin{figure}
	\centerline{\psfig{figure=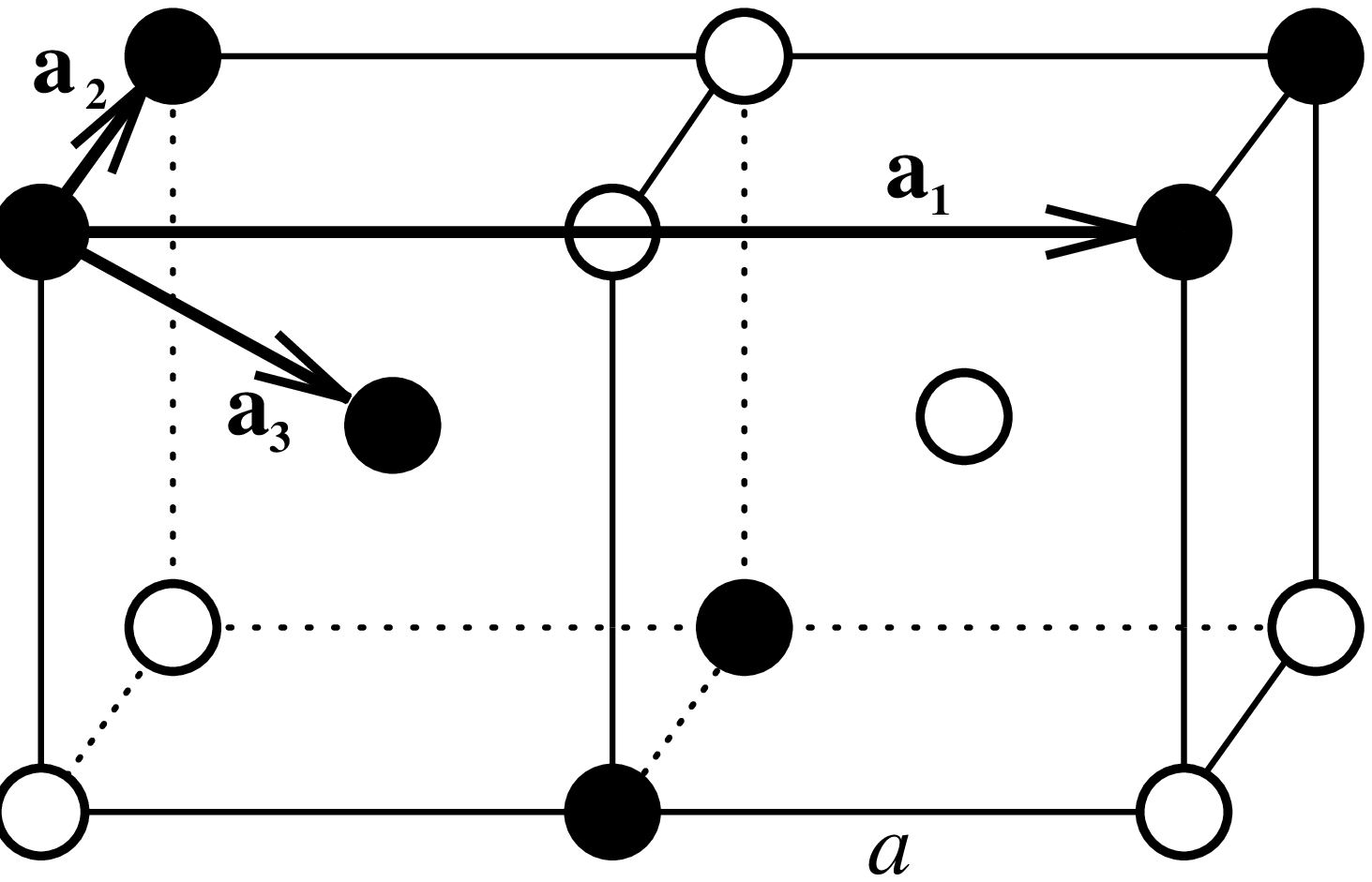,width=6cm,angle=270}
	\psfig{figure=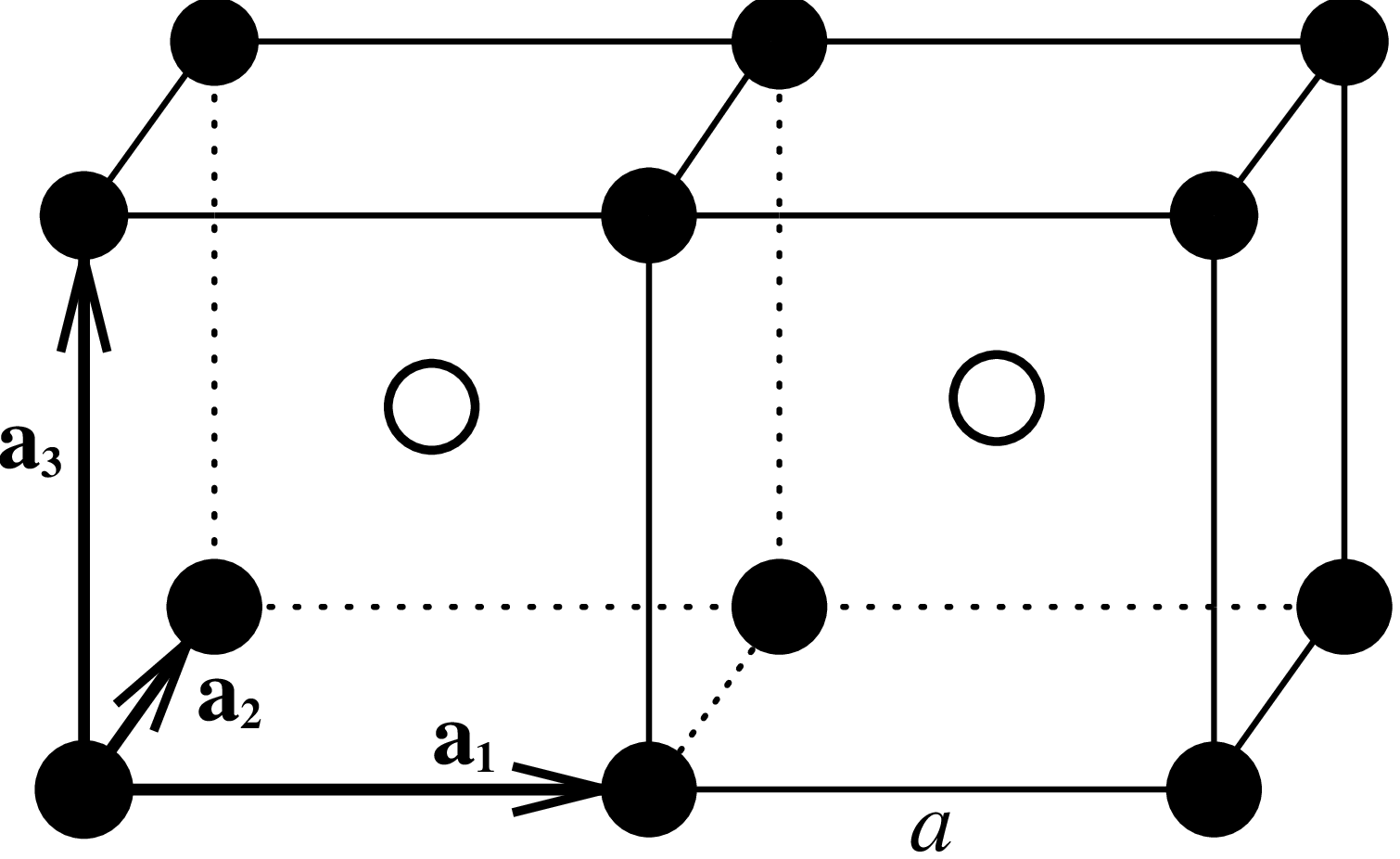,width=6cm,angle=270}}
      
      \centerline{{\Large AFM-(110) \hspace{20ex} AFM-(AB)}}
\caption{Decomposition of the bcc lattice into two equivalent 
sublattices. The AFM-(110)
configuration is built in such a way, that the (110)-planes alternately belong
to the sublattice $A$ and $B$. In the AFM-(AB) configuration all nearest
neighbours of a given lattice site belong to the other sublattice. The vectors
$\vec{a}_{1,2,3}$ are the translation vectors of the respective magnetic Bravais
lattice and $a$ is the lattice constant.\label{fig_afm_conf}}
\end{figure}

The sublattice density of states of the two antiferromagnetic configurations is
shown in Fig.\,\ref{fig_slqdos}. 
Contrary to the ferromagnetic
system, $\rho_{A\uparrow}(E)$ and $\rho_{A\downarrow}(E)$ occupy exactly the
same energy region due to the spin independence of the quasiparticle energies. 
The spectral weights which are connected to the quasiparticle energies show, 
however, a spin asymmetry leading to a spin dependent density of states. 
As for the paramagnetic and the ferromagnetic system there is the splitting
of the density of states into the lower and the higher energy region separated
by $U$.
In the AFM-(AB) phase an additional Slater gap occurs in the sublattice 
density of states
for all parameters,  
due to the high symmetry of the magnetic Brillouin zone. 
This  Slater gap is not visible in the
 AFM-(110) configuration.  
As the sublattice magnetization
$m_{A}$ increases, the shape of the sublattice density of states  in the
AFM-(110) phase becomes more and more similar to the density of states
of a two-dimensional system with sharp edges and a logarithmic singularity at
the center. This quasi two-dimensionality originates from the ferromagnetic
ordered  planes in the AFM-(110) configuration (see Fig.\,\ref{fig_afm_conf}).

\section{SUMMARY AND CONCLUSIONS}\label{sec_sum}
We presented a self-consistent moment approach to the Hubbard model which is
based on a two pole ansatz for the one electron spectral density.  A complete 
evaluation of this theory, which does not involve any further approximations,
leads to results identical to the method proposed by Roth. We believe, however,
the SDA to be physically more transparent and better motivated, due to the
analysis  in the strong coupling limit, which were essential to
formulate the two-pole ansatz. Since the SDA reproduces  
the  band limit, the atomic limit and the 
strong coupling limit, it is
expected to provide a reasonable interpolating solution  covering the
intermediate and strong coupling regime. This is confirmed in the paramagnetic
case  by the excellent agreement between the quasiparticle energies  calculated
with the SDA and recent Quantum Monte Carlo results \cite{BSW94a,BSW94b}. 
For the ferromagnetic case
we find a phase diagram very similar to the modified perturbation 
theory (SOPT-HF) \cite{BJ90} , which
does best for weak and intermediate Coulomb interaction.

From the presented phase diagram  we read off, that spontaneous ferromagnetism
exists only above a critical value of the band occupation and the Coulomb
interaction. Therefore, the SDA is much more restrictive than a simple mean
field theory (Stoner model). Antiferromagnetism is most stable for intermediate
coupling strength $U\approx W$ and in a small region around $n=1$. 
As a function of temperature the magnetization curves show  second order
phase transitions away from half filling. For band occupations close to 
half filling, however, the system exhibits first order transitions to the 
paramagnetic phase.
Note, that these first order transitions can not be observed by looking at the 
poles of the static paramagnetic susceptibility.
The corresponding
critical temperatures $T_{C}$ and $T_{N}$ lie for all parameters in a physically
reasonable region. All magnetic properties of the Hubbard model were shown to
find a direct explanation in the ($n$, $T$, $U$)-dependent QDOS.

By investigation of the system with and without the non-local term in the
self-energy, we find, that the inclusion of the $\vec{k}$-dependence leads  to
a narrowing of the quasiparticle subbands.  The correlation functions, which
represent the non-locality of the self-energy were discussed in detail. Here we
found, that almost all effect of the  $\vec{k}$-dependence is gathered in
the spin-spin correlation, which turns out to be negative, indicating,
therefore, antiferromagnetic correlation.  Consistent with the $d=\infty$
theory, the non-local terms rapidly become  less important as the dimension $d$
and the number of nearest neighbours $Z_{1}$ increases. 
For the three-dimensional bcc lattice, the influence of the 
$\vec{k}$-dependent terms in the self-energy  on the magnetic properties is
only marginal, whereas in two dimensions the corrections due to 
the non-locality
are essential. In two dimensions, however, we did restrict our investigations
to the paramagnetic case, since spontaneous magnetic order  in the
two-dimensional Hubbard model is excluded by the Mermin Wagner theorem 
\cite{MW66,Gho71}.
Very recent calculations indicate, that for the three-dimensional sc lattice
the influence of the non-local terms with respect to ferromagnetism is 
much more important than for the bcc lattice. 
We will discuss this fact in a forthcoming paper, where we investigate the 
influence of the lattice structure on ferromagnetism in the 
Hubbard model \cite{HN97}.

\acknowledgments{
This work has been done within the Sonderforschungsbereich 290 ("Metallische
d\"{u}nne Filme: Struktur, Magnetismus und elektronische Eigenschaften") of the
Deutsche Forschungsgemeinschaft.}

\newpage
\onecolumn


\befig{tp}
	\centerline{\psfig{figure=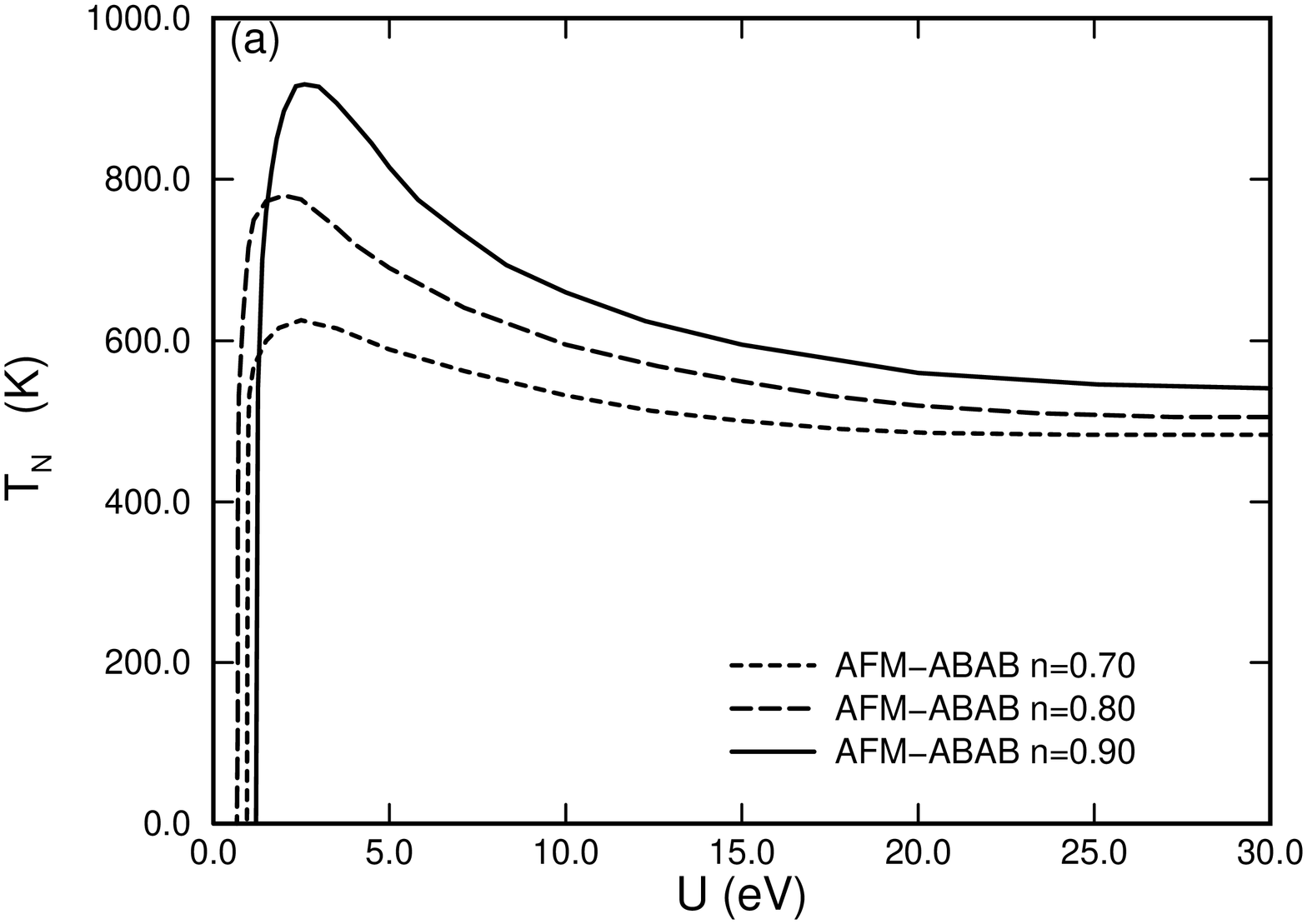,width=7cm,angle=270}
	\psfig{figure=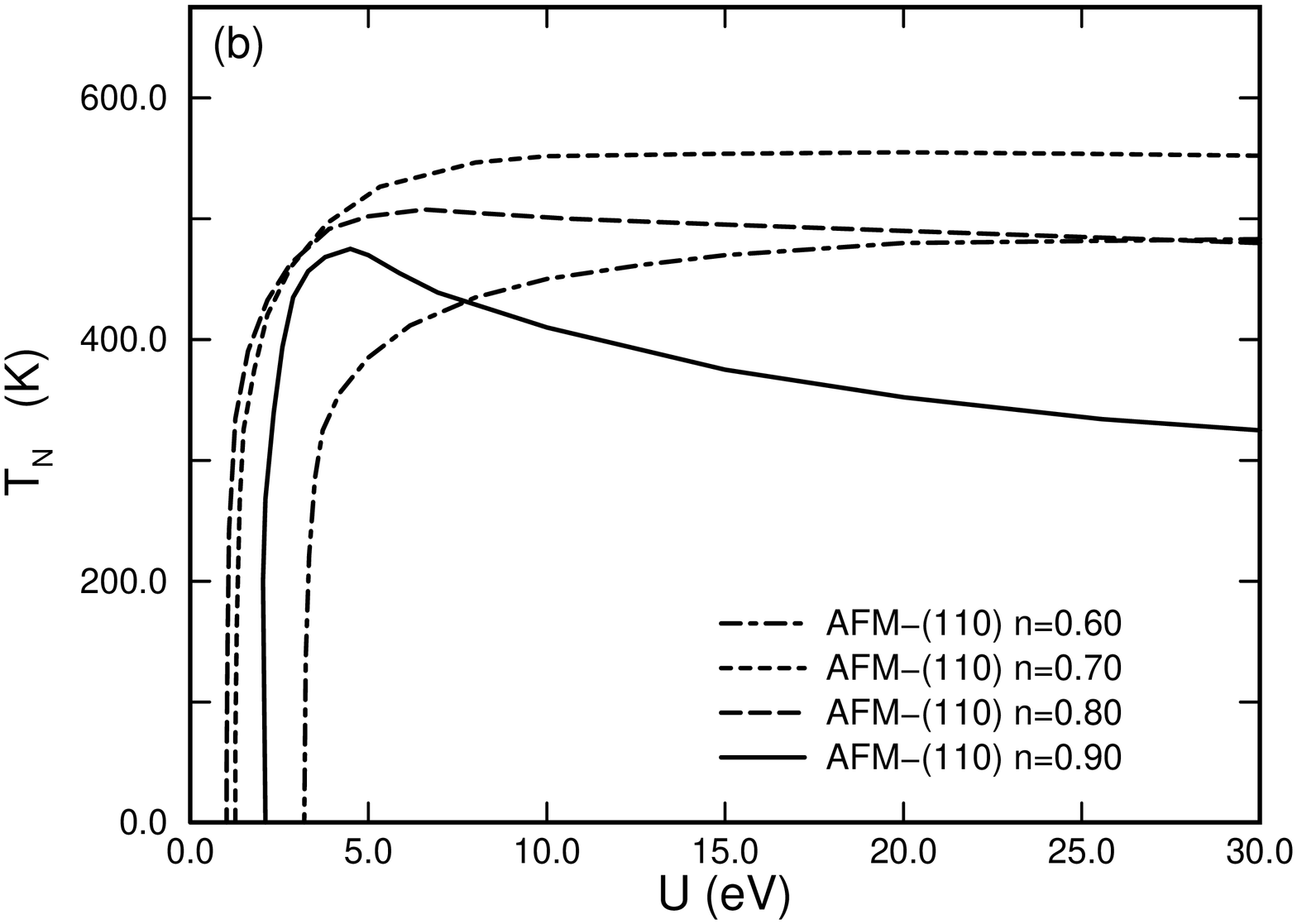,width=7cm,angle=270}}
	\caption{N\'{e}el-temperatures $T_{N}$ as a function of the Coulomb
	interaction $U$ for different band occupations  $n$;
	(a) AFM-(AB),  (b) AFM-(110). ($U=5.0\,$eV, $W=2.0\,$eV)
	\label{fig_Tn}}
\end{figure}

\befig{tp}
	\centerline{\psfig{figure=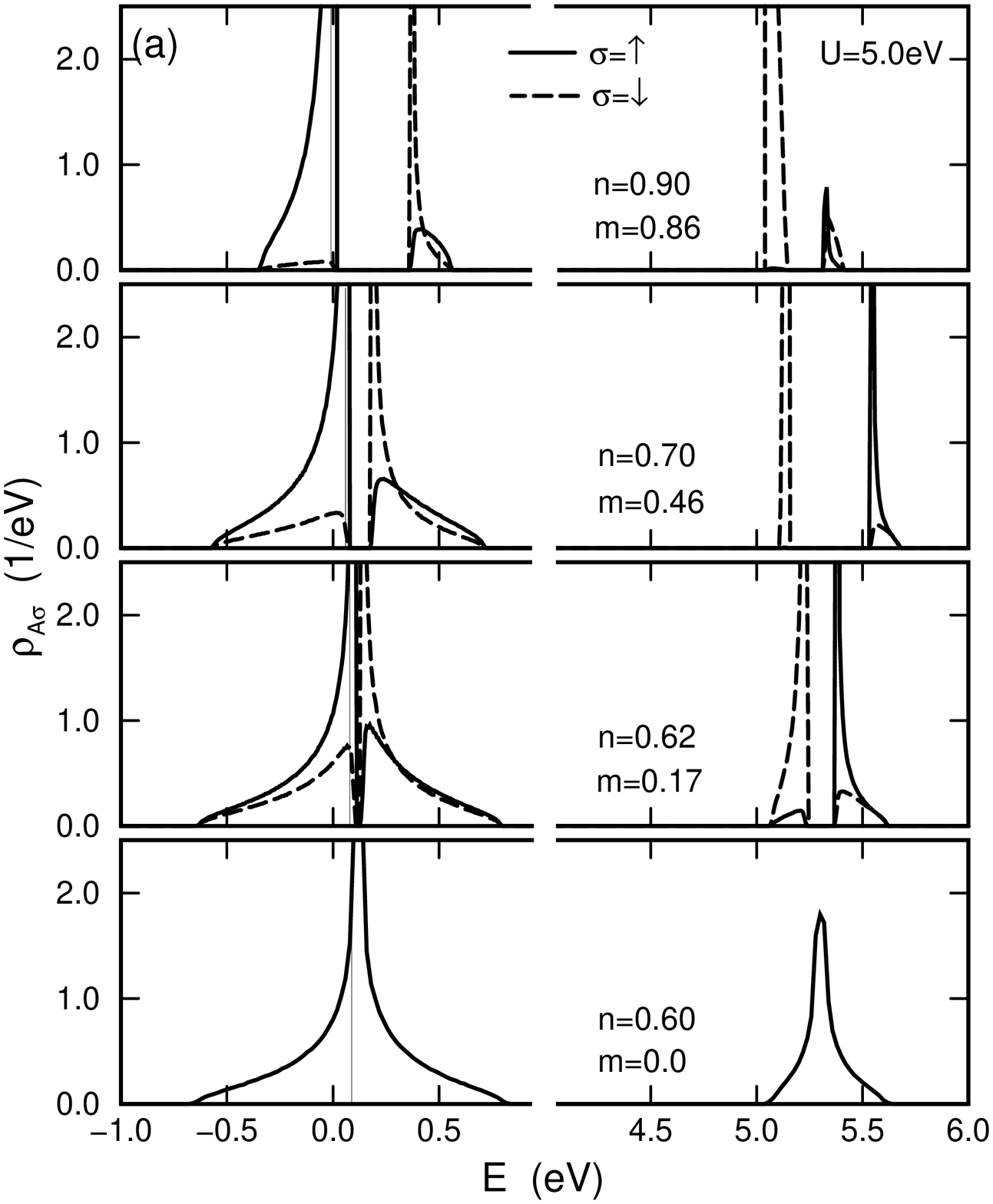,width=6cm,angle=0}
	\psfig{figure=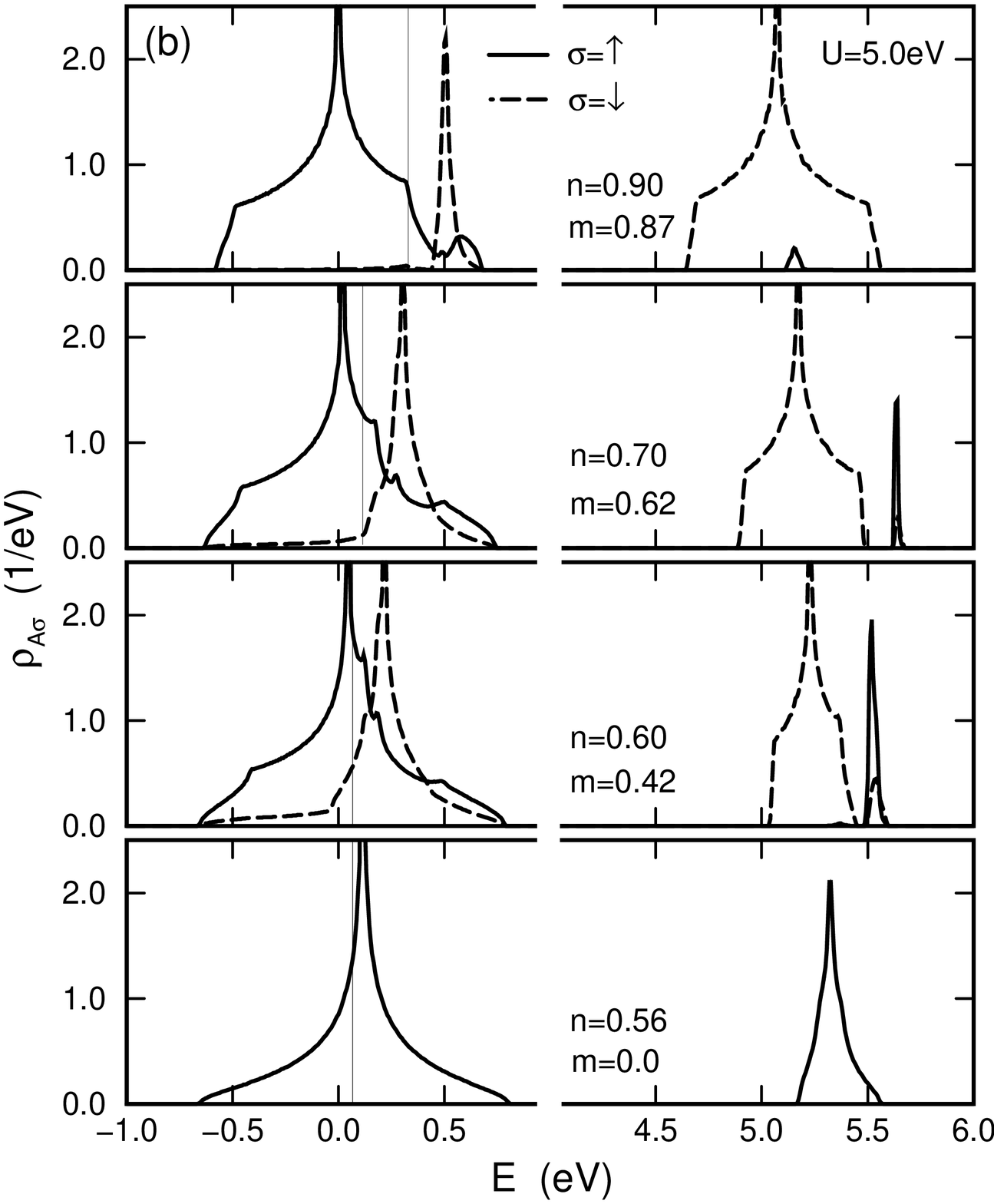,width=6cm,angle=0}}
	\caption{Sublattice density of states $\rho_{A\sigma}(E)$ for four
	different band occupations $n$; (a) AFM-(AB), (b) AFM(110).
	Solid lines correspond to the majority spin ($\sigma=\uparrow$), 
	broken lines to the minority spin direction ($\sigma=\downarrow$).
	The vertical lines indicate the positions of the chemical potential
	$\mu$.
	($W=2.0\,$eV, $T=0\,$K)\label{fig_slqdos}
	}
\end{figure}


\begin{thebibliography}{10}
\bibitem{foot1}
Besides ferromagnetic and commensurate antiferromagnetic order,
also incommensurate antiferromagnetic order is possible in the Hubbard model
(see for example \cite{FJ95}). Such calculations are not included here and are
left for future study.
\end{thebibliography}
\end{document}